\documentclass[12pt]{article}
\usepackage{latexsym}
\usepackage{amsmath}
\usepackage{tensor}
\usepackage{multibox}
\usepackage{verbatim}

\newcommand{\be}{\begin{equation}}
\newcommand{\ee}{\end{equation}}
\newcommand{\ba}{\begin{eqnarray}}
\newcommand{\ea}{\end{eqnarray}}
\def\vf{\varphi}

\def\cA{{\cal A}}
\def\cB{{\cal B}}
\begin{document}
\small
\thispagestyle{empty}

\renewcommand{\theequation}{\thesection.\arabic{equation}} \csname
@addtoreset\endcsname{equation}{section}

\begin{flushright}
hep-th/yymmddd \vskip 8pt {\today}
\end{flushright}

\vspace{5pt}

\begin{center}


{\Large\sc $(A)dS$ exchanges and partially-massless higher
spins}\\


\vspace{20pt}
{\sc D.~Francia${}^{a}$, J.~Mourad$^{b}$ and A.~Sagnotti$^{c}$}\\[15pt]

{${}^a$\it\sl\small Department of Fundamental Physics\\ Chalmers
University of Technology \\ S-412\ 96 \ G\"oteborg \ SWEDEN
\\ e-mail: {\small \it francia@chalmers.se}}\vspace{10pt}

{${}^b$\it\small AstroParticule et Cosmologie (APC) \footnote{Unit\'e mixte de Recherche du CNRS (UMR 7164)}\\
Universit\'e Paris VII - B\^atiment Condorcet \\
10, rue Alice Domon et L\'eonie Duquet \\ F-75205 Paris Cedex 13\
FRANCE
 \\ e-mail:
{\small \it mourad@apc.univ-paris7.fr}}\vspace{10pt}

{${}^c$\it\small
Scuola Normale Superiore and INFN\\
Piazza dei Cavalieri, 7\\I-56126 Pisa \ ITALY \\
e-mail: {\small \it sagnotti@sns.it}}

\vspace{10pt} {\sc\large Abstract}\end{center}

We determine the current exchange amplitudes for free totally
symmetric tensor fields $\vf_{\mu_1 \ldots \mu_s}$ of mass $M$ in a
$d$-dimensional $dS$ space,  extending the results previously
obtained for $s=2$ by other authors. Our construction is based on an
unconstrained formulation where both the higher-spin gauge fields
and the corresponding gauge parameters $\Lambda_{\mu_1 \ldots
\mu_{s-1}}$ are not subject to Fronsdal's trace constraints, but
compensator fields $\alpha_{\mu_1 \ldots \mu_{s-3}}$ are introduced
for $s > 2$. The free massive $dS$ equations can be fully determined
by a radial dimensional reduction from a $(d+1)$-dimensional
Minkowski space time, and lead for all spins to relatively handy
closed-form expressions for the exchange amplitudes, where the
external currents are conserved, both in $d$ and in $(d+1)$
dimensions, but are otherwise arbitrary. As for $s=2$, these
amplitudes are rational functions of $(ML)^2$,  where $L$ is the
$dS$ radius. In general they are related to the hypergeometric
functions $_3F_2(a,b,c;d,e;z)$, and their poles identify a subset of
the ``partially-massless" discrete states, selected by the condition
that the gauge transformations of the corresponding fields contain
some non-derivative terms. Corresponding results for $AdS$ spaces
can be obtained from these by a formal analytic continuation, while
the massless limit is smooth, with no van Dam-Veltman-Zakharov
discontinuity.

 \setcounter{page}{1}

\pagebreak

\tableofcontents

\section{Introduction}

In \cite{fms} we determined the current exchange amplitudes for
massless spin-$s$ totally symmetric higher-spin fields $\vf_{\mu_1
\ldots \mu_s}$ \cite{solvay} in a $d$-dimensional Minkowski space
time, and we also extended the analysis to massless fields in
$(A)dS$ backgrounds. In this analysis we resorted to an
unconstrained formulation for higher-spin fields, introduced and
developed in \cite{fs,st,fms}, where Fronsdal's trace constraints
\cite{fronsdal} on the gauge parameters $\Lambda_{\mu_1 \ldots
\mu_{s-1}}$ and the corresponding double-trace constraints on the
gauge fields are eliminated via the introduction, for $s \geq 3$, of
spin-$(s-3)$ compensators $\alpha_{\mu_1 \ldots \mu_{s-3}}$. This
unconstrained formalism has the peculiar feature of containing some
higher-derivative terms, and was motivated by the previous
unconstrained geometric non-local formalism of \cite{fs}. It is far
simpler, however, than the previous local constructions of
Buchbinder, Pashnev, Tsulaia and others \cite{bpt}, the first where
Fronsdal's constraints were overcome, starting from a BRST
\cite{BRST} construction, but at the price of introducing ${\cal
O}(s)$ additional fields. Further developments along these lines,
leading to an action with only a few additional fields, are
contained in \cite{buchnew}, while the extension of the non-local
formalism to the massive case and the reduction of the local
formalism of \cite{fs,st,fms} to two derivatives are described in
\cite{francia07}. The extension of the non-local geometric equations
to higher-spin fields of mixed symmetry is also available, and is
discussed in \cite{bb}. In this work we focus on the local
formulation for massive unconstrained higher-spin bosons in an
$(A)dS$ background. The corresponding analysis for the geometric,
non local massive theory, along the lines on \cite{francia07}, as
well as the extension to massive $(A)dS$ fermions, are interesting
issues that we leave for future work.

The key property of the external currents determining the exchanges
of \cite{fms} was the fact that they were conserved but otherwise
arbitrary, whereas in Fronsdal's formulation they would only need to
be partly conserved, while their double traces would be constrained
to vanish. Our flat-space result can be subsumed in the relatively
handy expression,
\be \label{fmsfin}
\langle J, \Box \, \vf \rangle = \sum_n \,  {1\over n!\, 2^{2n}(3-\zeta)_n}\ \langle J^{[n]}\;, \; J^{[n]}\rangle \, ,
\ee
that identifies the residue of the pole at the physical mass shell
of the exchanged higher-spin particle, where $J^{[n]}$ denotes the
$n$-th trace of the external current, the coefficient $\zeta$ is
defined as
\be
\zeta=\frac{d}{2}+s\, , \label{zeta}
\ee
$(a)_n$ denotes the $n$-th Pochhammer symbol of $a$, defined in
Appendix A, and the brackets denote a suitably normalized inner
product, so that
\be \langle \varphi , \varphi \rangle = \frac{1}{s!}
\ \varphi_{\mu_1 \ldots \mu_s}\, \varphi^{\mu_1 \ldots \mu_s}\, . \label{norm} \ee

In \cite{fms} the amplitudes of eq.~(\ref{fmsfin}) were shown to
propagate the correct numbers of on-shell degrees of freedom,
exactly like in Fronsdal's  formulation, even though the currents
involved were not doubly traceless. The analysis was also
generalized to the case of massless fields in maximally symmetric
$(A)dS$ space times, where a very similar result was obtained. The
main difference was the replacement of the flat-space D'Alembertian
with the Lichnerowicz operator \cite{lichn}, that in the index-free
notation of \cite{fms} is such that
\be
\Box_L\varphi =\Box_{AdS} \varphi+\frac{1}{L^2} \, \Big[ s(d+s-2) \varphi -2g\varphi' \Big] \, . \label{lichn} \ee
In this expression $\Box_{AdS}$ is the D'Alembertian for the $AdS$
background with metric $g$, $L$ is the $AdS$ radius and
$\varphi^\prime$ denotes the trace of $\varphi$. The Lichnerowicz
operator has the nice property of commuting with traces, covariant
derivatives and contractions, and the corresponding result for the
$dS$ background can be directly obtained from  the previous
expressions via the formal continuation of $L^2$ to $- L^2$.

The extension of these results to massive fields in Minkowski space
times is straightforward, essentially since the little groups for a
massive particle in $d$ dimensions and for a massless particle in
$(d+1)$ dimensions coincide. This simple correspondence underlies
the familiar description of massive $d$-dimensional fields as
Kaluza-Klein modes of massless $(d+1)$-dimensional fields. As a
result, the flat-space massive exchange can be simply obtained from
eq.~(\ref{fmsfin}) replacing $d$ with $(d+1)$ and taking into
account that the residue now refers to the physical mass shell of
the particle of mass $M$, and reads
\be \label{fmsfinm}
\langle J, (\Box - M^2) \, \vf \rangle = \sum_n \,  {1\over n!\, 2^{2n}\; \left(\frac{5}{2}-\zeta\right)_n}\ \langle J^{[n]}\;, \; J^{[n]}\rangle\ \, .
\ee
The key property of this residue is that its massless limit differs
markedly from the massless result of eq.~(\ref{fmsfin}). It should
be appreciated that the difference between the two expressions
generalizes to spin$-s$ fields the van Dam-Veltman-Zakharov
(\emph{vDVZ}) discontinuity originally found in \cite{vdvz} for the
spin-two exchange in Minkowski space.

A naive attempt to extend this result to $(A)dS$ backgrounds would
readily run into difficulties. This can somehow be anticipated from
the very nature of the corresponding unitary representations
\cite{rep1,rep2}: aside from massive fields, describing states of
the continuous series, and massless fields, whose degrees of freedom
are maximally reduced, these backgrounds allow a number of
intermediate options, usually termed \emph{partially massless
fields} \cite{partlymass}. The corresponding states belong to the
discrete series of unitary $SO(1,d)$ representations, whose degrees
of freedom are intermediate, in number, between those of massless
and massive fields. For a spin-$s$ field there are $s-1$ special
values for the mass $M$. They all lie below the lower end
$(ML)^2=(s-1)(d+s-4)$ of the continuous series, and feature the
emergence of surprising gauge symmetries. The corresponding
parameters have spins $0,1,\ldots, s-2$, all of which are smaller
than $s-1$, the spin of the more familiar massless gauge parameters
$\Lambda_{\mu_1 \ldots \mu_{s-1}}$. For one matter, discrete states
should somehow show up in the current exchanges, as they do for
$s=2$ \cite{higuchi,porrati}, and this explains the relative
complexity of these quantities in $(A)dS$ backgrounds.
Alternatively, one can observe, in analogy with the known $s=2$
result, that the exchange amplitudes depend on $(ML)^2$, and exhibit
two distinct limiting behaviors in the flat massless case,
$(ML)^2\to 0$, and in the flat massive case, $(ML)^2 \to \infty$. As
such, in accord with Liouville's theorem, they must have zeros and
poles, and the latter, as we shall see, can be associated in general
to a subset of the partially massless points.

As is well known, de Sitter space times can be represented as
quadratic surfaces in higher-dimensional Minkowski space times, and
this suggests to perform a ``polar'' decomposition, much along the
lines of what is usually done for spheres in a Euclidean space.
Hence, a massless field in a $(d+1)$-dimensional Minkowski space
time with a suitable ``radial'' dependence should give rise to an
effective massive $dS$ field. This interesting correspondence,
discussed in \cite{fronads} and recently reconsidered in
\cite{bsieg}, will be reviewed in Section 3 \footnote{Alternative
treatments of massive totally symmetric $(A)dS$ fields can be found
in \cite{Zinoviev:2001dt}.}. As a byproduct, the procedure will also
establish a neat correspondence between discrete states and
partially massless fields with residual gauge symmetries. The
unconstrained formulation of \cite{fs,st,fms} is particulary
convenient in this context, since for generic values of the mass $M$
it allows a gauge choice eliminating all tensor fields with at least
one radial index. The remaining difficulty is related to the
compensator fields, that do not vanish in the presence of an
external current and must be disentangled, while a key novelty with
respect to the flat case is that a conserved current must have a
well-defined pattern of non-vanishing radial components. This is due
to the combination of two facts, the conservation equation in the
$(d+1)$-dimensional Minkowski space and the condition that the $dS$
current be covariantly conserved. Notice, however, that the first
property is forced upon by the construction while the second is not
inevitable, but is added here in analogy with what was done for the
massive flat-space construction in \cite{fms}. With this proviso,
the full current is constructed in Section 4, while its relation to
the compensators is spelled out in Section 5. The intermediate
expressions will be somewhat lengthy, but the final result for the
current exchanges, as we shall see in Section 7, can be presented in
a rather compact closed form, where the $(3-\zeta)_n$ Pochhammer
symbols of the flat-space expression (\ref{fmsfin}) are replaced by
interesting analytic functions of $(ML)^2$, that can be related to
the generalized hypergeometric function $_3F_2(a,b,c;d,e;z)$. The
final expressions are rational functions of $(ML)^2$, although this
property is hardly evident at first sight, with poles that identify
a \emph{subset} of the discrete states. These particular discrete
states are selected because the gauge invariance of the
corresponding Lagrangians would require that the external currents
satisfy conservation conditions involving some of their own traces.
In Section 6 we shall also discuss the massless limit of the $dS$
exchanges, that strictly speaking should be taken for the $AdS$
case, after a suitable analytic continuation, not to cross the
unitarity gap present in $dS$ space times. As we shall see, the
result obtained in \cite{higuchi,porrati}, namely that for
linearized gravity the $vDVZ$ discontinuity disappears in the
presence of a cosmological constant, generalizes to all higher-spin
cases. Let us stress that our results concern free fields: the
arguments of \cite{duff}, that ascribe the disappearance of the
discontinuity for $s=2$ to the classical approximation, or the
considerations of \cite{porrati2}, are clearly out of reach at the
present time for higher spins. With this proviso, our conclusion
will be that the flat limit of all spin-$s$ $AdS$ current exchanges
is smooth, with no $vDVZ$ discontinuity, and coincides with the
massless flat-space result of \cite{fms}. We end in Section 8 with a
brief discussion of our results, while the three Appendices fill
some technical gaps of our derivations.

\section{Massive current exchanges in Minkowski space times}

In this Section we review how the massless current exchanges of
\cite{fms} also encode the relevant information on massive exchanges
in Minkowski space times. This discussion is also meant to review
briefly the unconstrained formulation of \cite{fs,st,fms} and to
stress its advantages for obtaining explicit expressions for the
current exchanges.

\subsection{Massive spin-$s$ fields from massless $(d+1)$-dimensional fields}

Our starting point is the Fronsdal equation,
\be {\cal F}_{\mu_1 \ldots \mu_s} \equiv \Box \varphi_{\mu_1 \ldots
\mu_s} - \left( \partial_{\mu_1} \partial \cdot \varphi_{\mu_2
\ldots  \mu_s} + \ldots \right) + \left( \partial_{\mu_1}
\partial_{\mu_2} \varphi^\prime_{\mu_3 \ldots \mu_s}\right) \, = \,
0  \, , \label{fronsdeq} \ee
where the omitted terms in the ``Fronsdal operator" ${\cal F}$ complete the symmetrization
in the $s$ indices $\mu_1$, $\ldots$, $\mu_s$ and the ``prime'' denotes a trace. The key feature
of this expression is that it \emph{is not} fully gauge invariant, but rather transforms as
\be
\delta {\cal F}_{\mu_1 \ldots \mu_s} = \frac{1}{2}\ \partial_{\mu_1} \partial_{\mu_2} \partial_{\mu_3}
\Lambda^\prime_{\mu_4 \ldots \mu_s} + \ldots \, ,
\ee
where the remaining, omitted terms, are $s! -1$, under the natural
gauge transformation
\be \delta \varphi_{\mu_1  \ldots \mu_s} = \partial_{\mu_1}
\Lambda_{\mu_2  \ldots \mu_s} + \ldots \, . \label{gauges} \ee

In the compact, index-free notation of \cite{fs,st,fms}, these expressions read simply
\be {\cal F} = \Box \varphi - \partial \, \partial \cdot \varphi +
\partial^2 \varphi^\prime \, , \qquad \delta \varphi = \partial
\Lambda  \, , \ee
and
\be \delta {\cal F} = 3 \, \partial^3 \Lambda^\prime \, . \ee
For $s \geq 3$, in \cite{fs,st} we thus introduced a compensator $\alpha_{\mu_1 \ldots \mu_{s-3}}$,
such that
\be \delta \alpha = \Lambda^\prime \, , \ee
and in \cite{fms} we combined ${\cal F}$ and $\alpha$ into the
gauge-invariant tensor
\be {\cal A} = {\cal F} - 3 \, \partial^3 \, \alpha \, . \ee

In this index-free notation, the general gauge invariant flat-space Lagrangian takes the relatively
simple form
\be \label{boselagr} {\cal L} \, = \, \frac{1}{2} \, \vf \, \left(
\cA \, - \, \frac{1}{2} \ \eta \, \cA^{\; \prime} \right) \, - \,
\frac{3}{4} \ {s \choose 3 } \, \alpha\, \partial \cdot \cA^{\;
\prime} \, + \, 3 \, { s \choose 4 } \, \beta \, \left[ \vf^{\;
\prime\prime} \, - \, 4 \, \partial \cdot \alpha \, - \, \partial \,
\alpha^{\, \prime} \right]\,- \, J \; \vf \, , \ee
where the coefficients are expressed in terms of the binomial
coefficients
\be
\left( n \atop k \right) = \frac{n!}{k!\; (n-k)!} \, . \ee
We have also added, for future convenience, an external current $J$,
which is to be \emph{conserved} in order that ${\cal L}$  be gauge
invariant up to total derivatives. Notice that, aside from the gauge
field $\vf$ and the compensator $\alpha$, this Lagrangian also
involves a spin-$(s-4)$ Lagrange multiplier $\beta$, that first
presents itself for $s=4$ and whose gauge transformation is
\be \label{transbeta} \delta \beta \, = \, \partial \cdot \partial \cdot \partial \cdot \Lambda \,
.\ee
The resulting field equation for $\vf$ is
\be {\cal A} \ - \ \frac{1}{2} \ \eta \ {\cal A}^{\; \prime} \ + \eta^2 \ {\cal B}
= \ J \, , \label{eqaJ} \ee
where
\be {\cal B} \, \equiv \, \beta \, - \, \frac{1}{2} \, (\partial \cdot \partial \cdot \vf^{\, \prime} \, -
\, 2 \,  \Box \, \partial \cdot \alpha \, - \, \partial \, \partial \cdot \partial \cdot \alpha) \, , \ee
is the gauge-invariant completion of the Lagrange multiplier
$\beta$.

In the absence of an external current, eq.~(\ref{eqaJ}) can be turned into
\be
{\cal A} = 0 \, , \label{eqasimple}
\ee
and eventually into the Fronsdal equation (\ref{fronsdeq}), after a
partial gauge fixing using the trace $\Lambda^{\; \prime}$ of the
gauge parameter. On the other hand, the field equation for $\beta$
replaces,  in this formulation, Fronsdal's double trace constraint,
and relates the double trace of $\vf$ to the compensator $\alpha$
according to
\be
\vf^{\; \prime\prime} = 4 \partial \cdot \alpha + \partial \alpha^{\;\prime} \, . \label{nodtr}
\ee
One can also show that the double trace of ${\cal A}$ vanishes
identically if one makes use of eq.~(\ref{nodtr}) .

Let us now reconsider these results in a slightly different
notation, to which we shall resort in this paper since it will prove
particularly convenient for some of our derivations. Let us
therefore introduce a \emph{constant} auxiliary vector, $u^\mu$, so
as to replace the fully symmetric tensor $\varphi_{\mu_1 \ldots
\mu_s}$ with the index-free expression
\be \varphi(x,u^\mu)\ =\ {1\over s!}\
\varphi_{\mu_1\dots\,\mu_s}\,u^{\mu_1}\dots u^{\mu_s} \ .
\label{hdef} \ee
The Fronsdal equation (\ref{fronsdeq}) then takes the form
\footnote{Here $u \cdot \partial$ computes the gradient, $\partial
\cdot  \partial_u$ the divergence and $\partial_u \cdot
\partial_u$ the trace.}
\be {\cal F}(x,u^\mu) \equiv \left[\Box\ -\ (u\cdot
\partial)\left(\partial\cdot {\partial_u}\right)\ +
\ {1\over 2}\ (u\cdot\partial)^2 \left({ \partial_u}\cdot{
\partial_u}\right)\right]\varphi(x,u^\mu)\ =\ 0\, ,\label{froo} \ee
while the corresponding equation (\ref{eqasimple}) of the unconstrained formalism reads
\be \left[\Box\ -\ (u\cdot
\partial)\left(\partial\cdot {\partial_u}\right)\ +
\ {1\over 2}\ (u\cdot\partial)^2 \left({ \partial_u}\cdot{
\partial_u}\right)\right]\varphi(x,u^\mu)\, -\, \frac{1}{2} \ \left(u \cdot \partial \right)^3 \,
\alpha(x,u^\mu) \,= \, 0\, . \label{frooa} \ee

One more specification is needed, since our aim here is connecting
massive $d$-dimensional higher-spin fields to massless fields in a
$(d+1)$-dimensional Minkowski space time. The starting point is thus
a spin-$s$ field in a $(d+1)$-dimensional Minkowski space time, with
a corresponding $(d+1)$-dimensional auxiliary vector $U^A$, or
equivalently with two types of auxiliary variables, a
$d$-dimensional vector $u^\mu$ and a scalar $v$, the internal
component of $U^A$ along the extra dimension. The dependence on the
additional Kaluza-Klein coordinate $y$ is simply chosen to be
$e^{iMy}$, with $M$ the resulting mass in $d$ dimensions, so that
one is led to the expansions
\ba && \vf(X,U^A)\equiv
e^{iMy}\vf(x,U^A)=e^{iMy}\sum_{r=0}^{s}{1\over
r!}\ \vf_{s-r}(x,u^\mu)\, v^{r}\, , \nonumber \\
&& \Lambda(X,U^A)\equiv
e^{iMy}\Lambda(x,U^A)=e^{iMy}\sum_{r=0}^{s-1}{1\over r!}\
\Lambda_{s-1-r}(x,u^\mu)\, v^{r}\, ,\label{frooa2} \ea
for the field $\vf$ and the gauge parameter $\Lambda$, where $\vf_r$
and $\Lambda_r$ denote $d$-dimensional spin-$r$ quantities. In the
following we shall factor out the $y$ dependence whenever possible,
for brevity. In this notation, the gauge transformation in $(d+1)$
dimensions reads
\be \delta \vf(x,U^A)=u \cdot \partial\Lambda(x,U^A)+i v
M\Lambda(x,U^A) \, , \ee
that in terms of the $d$--dimensional component fields becomes
\be \delta \vf_r(x)= \partial\Lambda_{r-1}(x)+i{M}
(s-r)\Lambda_{r}(x), \quad (r=0,\dots s) \, . \label{gauge} \ee

In a similar fashion, the spin-$(s-3)$ compensator $\alpha_{A_1
\ldots A_{s-3}}$ present in $(d+1)$ dimensions transforms as
\be \delta \alpha(x,U^A)=(\partial_u \cdot
\partial_u+\partial^2_v)\, \Lambda(x,U^A) \, , \ee
that in terms of $d$-dimensional components becomes
\be \delta \alpha_r(x)= \Lambda^\prime_{r+2}(x)+\Lambda_{r}(x)\, ,
\quad (r=0,\dots s-3)\, . \ee
The two fields $\vf$ and $\alpha$ determine, as we have seen, the gauge invariant modification ${\cal A}$ of the Fronsdal operator, that in this notation reads
\be \cA(x,U^A) \, =\, {\cal F}(x,U^A)- \frac{1}{2}\ (U \cdot
\partial)^3\alpha(x,U^A) \, . \ee
As we have seen, in this unconstrained formalism the double trace of
$\vf$ does not vanish identically, but can be fully expressed in
terms of the compensator. Moreover, as we have stressed the double
trace $\cA^{\prime\prime}$ of ${\cal A}$ vanishes identically after
using eq.~(\ref{nodtr}).

Notice that the gauge transformation of eq.~(\ref{gauge}) involves
shifts that clearly allow the elimination of the Stueckelberg modes
$\vf_r$ with $r=0,\dots s-1$. In the resulting gauge
$\partial_v\vf(x,U^A)=0$, so that the free equations reduce to
\be
(\Box-M^2)\vf-(u\cdot \partial+iMv)\; \partial \cdot \partial_u\vf
+{1\over 2}\, (u \cdot \partial +iMv)^2\partial_u \cdot
\partial_u\vf-{1\over 2}\, (u \cdot \partial+iMv)^3\alpha=0 \, . \ee
This determines, in particular, the compensator field, whose
components $\alpha_0, \ldots, \alpha_{s-3}$ all vanish, as can be
seen considering, recursively, the coefficients of $v^s$ for $s \geq
3$. The coefficient of $v^2$ then gives the trace condition
\be \label{fp1} \partial_u \cdot \partial_u\; \vf=0 \, ,\ee
while the coefficient of $v$ gives the divergence condition
\be \label{fp2}
\partial \cdot \partial_u\; \vf=0 \, .
\ee
Finally the terms independent of $v$ give the mass shell condition
\be \label{fp3}
(\Box-M^2)\; \vf=0\, .
\ee
In other words, one thus recovers the Fierz-Pauli conditions,
summarized in eqs.~(\ref{fp1})-(\ref{fp3}), that identify a
traceless and divergence-free tensor $\vf_{\mu_1 \ldots \mu_s}$
describing an irreducible set of massive spin-$s$ modes in a
$d$-dimensional Minkowski space time.

\subsection{Coupling to an external current and the vDVZ discontinuity}

We can now turn to a brief review of the coupling between
$(d+1)$-dimensional massless fields and conserved external currents.
In the unconstrained formulation of \cite{fs,st,fms}, the relevant
equations in $(d+1)$ dimensions are
\be \cA \ - \ \frac{1}{2} \ \eta \ \cA^{\; \prime} \ + \eta^2 \ \cB
= \ J \, .\label{eqaJu} \ee
The condition $\cA^{\prime\prime}=0$ still holds, after making use
of eq.~(\ref{nodtr}), so that taking successive traces
eq.~(\ref{eqaJu}) can be turned into
\be \langle J, \cA  \rangle=\sum_{n=0}^{N}\ \rho_n(d-1,s){1 \ \over
{n!\; \; 2^n}}\ \langle J^{[n]}, J^{[n]}\rangle \, , \ee
where $\langle\varphi,\varphi\rangle$ is defined in
eq.~(\ref{norm}).

In \cite{fms} the coefficients $\rho_n(d-1,s)$ were related to a
difference equation, whose solution can be cast in the form
\be \rho_n(d-1,s)={1\over 2^n\left({5\over
2}-\zeta\right)_n}\, , \ee
with $(a)_n$ the $n$-th Pochhammer symbol of $a$, defined in
Appendix A, and $\zeta$ is defined in eq.~(\ref{zeta}).

Let us now choose again for the dependence on the extra dimension of
fields and currents the exponential $e^{iMy}$, where $M$ will be the
resulting mass for the spin-$s$ field. Current conservation in
$(d+1)$ dimensions then reads
\be
\partial_v J={i\over M}\ \partial \cdot \partial_u J\, ,
\ee
so that, integrating this equation, the $v$ dependence is fully
encoded in
\be
J(x,U^A)=e^{{iv\over M}\partial \cdot \partial_u}
J_s(x,u^\mu)\label{curr}\, . \ee
Let us also add the condition that the $d$-dimensional current be conserved, so that
\be
\partial \cdot \partial_u J_s(x,u^\mu)=0 \, .
\ee
Strictly speaking, this further condition would appear not fully
motivated for massive fields, but by explicitly manipulating the
field equations for the first few low-spin examples one can convince
oneself that, even starting from elementary non-conserved currents,
the field equations can be conveniently recast in a form such that
the exchanges involve effectively conserved currents. For instance,
for spin $s=1$ the Maxwell-Proca equation reads
\be \label{s1eq}
\Box A_\mu - \partial_\mu \partial \cdot A - M^2 A_\mu = J_\mu \, ,
\ee
whose divergence implies that
\be
\partial \cdot A = - \, \frac{1}{M^2}\ \partial \cdot J \, ,
\ee
so that eq.~(\ref{s1eq})  can be turned into
\be \left( \Box - M^2 \right) A_\mu = \tilde{J}_\mu  \, ,
\label{massivepole} \ee
where the effective current is
\be \tilde{J}_\mu = J_\mu - \frac{1}{M^2} \ \partial_\mu\, \partial
\cdot J \, . \ee
 Notice that, on shell, the effective
current $\tilde{J}_\mu$, defined as the residue at the physical
pole, is actually conserved, as advertised, while working from the
beginning with a conserved current has the effect of hiding the
singularity of this expression at $M=0$, and thus allows for a
 massless limit that is clearly smooth.

Once one makes the choice of working with conserved currents, in
flat space the complete current does not depend on $v$ or,
equivalently, all of its components with indices along the
additional dimension vanish, so that $J$ reduces to $J_s$. The
contraction of the spin-$s$ field equation (\ref{eqaJu}) with a
conserved current then gives
\be \label{massiveflat} \langle J, (\Box-M^2)\vf \rangle=
\sum_{n=0}^{N}\ \rho_n(d-1,s) \ {1\over {n!\; \; 2^n}}\ \langle
J_s^{[n]}\cdot J_s^{[n]}\rangle \, . \ee
As we anticipated, the massless limit of this expression is indeed
regular, but does not coincide with the massless current exchange,
that has a similar form but for the key replacement of
$\rho_n(d-1,s)$ with $\rho_n(d-2,s)$. It is important to stress that
the existence of a non singular massless limit is guaranteed by the
conservation of the $d$-dimensional current: without this crucial
condition, the limit would be singular, as can be foreseen from
eq.~(\ref{curr}), or from eq.~(\ref{massivepole}).

We can end this section by showing how the form of the current
exchange amplitude of eq.~(\ref{massiveflat}) makes unitarity
manifest. In general, the issue at stake is whether the residue of
the propagator pole is positive for the generic conserved currents
allowed in the exchange. The key observation to this effect is that,
due to current conservation, only the spatial components of the
current enter the contractions, giving rise to terms of the type
\be \tilde J_{\
a_1\dots a_{s}}\tilde J^{a_1\dots a_{s}} \, , \label{exchmassl}
\ee
where the $a_i$ indices are \emph{transverse} to the momentum, that
is time like for $M^2 >0$, and $\tilde J$ is a \emph{traceless} and
\emph{conserved} current, determined by the projection of
eq.~(\ref{massiveflat}). Let us stress that this argument actually
proves the positivity of the current exchange \emph{only} for
$M^2>0$, since only in this case is the on-shell momentum time like,
while for $M^2<0$ it would be space like. One thus recovers a
well-known fact: both the positivity of current exchanges in
Minkowski space and the unitarity of the coupling require a
\emph{positive} squared mass for all spins $s\geq 1$, although the
restriction on the sign of $M^2$ is not necessary for $s=0$.

\section{Massive $(A)dS$ fields via a radial dimensional reduction}

Let us now consider a curved space-time, with $e^a(x)= e_\mu^a(x) \,
dx^\mu$ a moving basis, where $e_\mu^a(x)$ is the vielbein, and let
$e^\mu_a(x)\, {\partial_\mu}$ be the dual vector fields, where
$e^\mu_a(x)$ is the inverse vielbein, such that
\be
e_\mu^a(x) \ e^\mu_b(x) = \delta^a_b \, .
\ee
Under an infinitesimal Lorentz transformation
\be \delta e^a = \epsilon^a{}_b(x) \, e^b(x) \, , \label{lorentz}
\ee
the spin connection $\omega^{ab}(x)$ transforms as
\be
 \delta \omega^{ab}(x) =\epsilon^{a}{}_{c}(x) \, \omega^{cb}(x)
 +\epsilon^b{}_{c}(x)\, \omega^{ac}(x)-d\epsilon^{ab}(x)\, ,
\ee
so that the corresponding curvature two-form is
\be
 R^{ab}=d\omega^{ab}+\omega^{ac}\wedge \omega_{c}{}^{b}\, ,
\ee
with
\be R^{ab}=\frac{1}{2}\ R_{\mu\nu}{}^{ab} dx^\mu \wedge dx^\nu \, .\ee

In the moving basis, a rank-$s$ fully symmetric tensor field
takes the form
\be \vf(x)={1\over s!}\ \vf_{a_1\dots a_s}(x)\, e^{a_1}(x) \otimes
\dots \otimes\, e^{a_s}(x) \, . \ee As in the previous section, let
us also introduce a \emph{fixed} and \emph{coordinate-independent}
auxiliary vector $u^a$, or, better, a collection of auxiliary
constants, in order to associate to the tensor a function
$\vf(x,u^a)$ that is homogeneous of degree $s$ in $u$:
\be \vf(x,u^a)={1\over s!}\ \vf_{a_1\dots a_s}(x)\, u^{a_1}\dots
u^{a_s}\, . \ee
Under the infinitesimal local Lorentz transformation (\ref{lorentz})
the tensor components also rotate, and the same is true for the
composite function $\vf(x,u^a)$, since $u^a$ is a fixed vector.
Nonetheless, the effect of the transformation can be conveniently
mimicked by
\be \delta \vf(x,u^a)={1 \over 2}\ {\epsilon^{ab}S_{ab}}\,
\vf(x,u^a) \, , \ee
with
\be
S_{ab}=u_a {\partial \over \partial u^b}-
u_b {\partial \over \partial u^a} \equiv  u_a \partial_{u^b}-
u_b \partial_{u^a} \, .
\ee
This is the starting point for a convenient algorithm to derive all
the subsequent results. For instance, it leads directly to define
the covariant derivative
\be D_\mu\vf(x,u^a)=\left(\partial_\mu+{1 \over 2}\
{\omega_\mu^{ab}\; S_{ab}}\right)\, \vf(x,u^a)\, . \label{covdev}
\ee

In our applications to $(A)dS$ space times the commutator of a pair
of covariant derivatives (\ref{covdev}) will thus recover the
curvature tensor
\be [D_\mu,D_\nu]\chi(x,u^a)={1 \over 2}\
R_{\mu\nu}{}^{ab}{S_{ab}}\; \chi(x,u^a) \, , \ee
since in Einstein's gravity the torsion tensor vanishes identically.

Let us now define
\be e_a^\mu\; D_\mu=D_a \, , \qquad {\rm and} \qquad D=u^a\; D_a\, .\ee
Just like the function $\vf(x,u^a)$ is associated to the tensor
$\vf_{\mu_1 \ldots \mu_s}$, the function $D\vf$ is then associated
to the tensor whose components in the moving basis are the
symmetrized gradient of $\vf_{\mu_1 \ldots \mu_s}$,
\be D_{a_1}\vf_{a_2\dots
a_{s+1}}+D_{a_2}\vf_{a_1\dots a_{s+1}}+\dots \,  . \ee
Notice that, with respect to the natural scalar product
\be
\int d^d x \, det \; e \, \ \langle\vf ,\vf \rangle \, ,\label{scalarflat}
\ee
where
\be
\langle\vf ,\vf \rangle={1\over s!} \
 \vf_{a_1\dots a_s}\, \vf^{a_1\dots a_s}\, , \ee
the adjoint
of $D$ is
\be D^\dagger=-\, \partial_{u^a} \,  D_a \, .\label{ddag} \ee
We can conclude this section by displaying the useful commutator
\be [D,D^\dagger]=D^aD_a-{1\over
4}\ R_{ab}{}^{cd}\, S^{ab}\, S_{cd}\, . \label{commut} \ee
As we shall see, this notation will prove particularly useful in the following sections.

\subsection{Foliating a flat $(d+1)$-dimensional space time by $dS$ sections}

Let us now consider a $(d+1)$-dimensional Minkowski space time with coordinates
$X^{\hat \mu}$, and let us foliate it by ``de Sitter" sections with constant
$R^2=X^{\hat \mu}X^{\hat\nu}\eta_{\hat \mu\hat \nu}$ \cite{bsieg}. If the
$dS$ coordinates are denoted by $x^\mu$, the flat metric can thus be presented in the ``polar'' decomposition
\be
ds^2=dR^2+R^{2}\, g_{\mu\nu}(x)\, dx^\mu dx^\nu, \label{metricR} \ee
where
$g_{\mu\nu}(x)$ is the metric for a $dS$ space of unit radius. A more
convenient parametrization obtains letting
\be
z=L \log\left(\frac{R}{L}\right)\, ,
\ee
with $L$ a length scale to be identified with the $dS$ radius, since
eq.~(\ref{metricR}) then becomes
\be \label{metricz}
ds^2=e^{2z\over L}(dz^2+ds^2_{dS})\, , \ee
where
\be
ds^2_{dS} = L^2 \ g_{\mu\nu}(x)\, dx^\mu dx^\nu
\ee
is the metric for a $dS$ space of radius $L$. Notice that
this foliation actually covers half of the original Minkowski space time, the region with $X^{\hat \mu}X_{\hat \mu}>0$, where the radial coordinate, or the corresponding $z$, are real.
Notice also that the $z=-\infty$ hypersurface corresponds to the light cone.

A moving frame for the
$(d+1)$-dimensional flat space is then given by
\be
{\tilde e}^A=({\tilde e}^{z},{\tilde e}^a) = e^{z\over L}\ ({e}^{z},{e}^a) \, ,\ee
where $e^z$ is simply $dz$ and $e^a$ is a $dS$ moving frame. As a result, the complete spin connection for the $(d+1)$-dimensional flat space decomposes as
\be
{\tilde \omega}^{ab}=\omega^{ab},\qquad {\tilde \omega}^{az}=\frac{1}{L}\ e^{\;- {z\over
L}}\ e^a\, ,
\ee
where $\omega^{ab}$ is the corresponding $dS$ spin connection.

A spin-$s$ field in the $(d+1)$-dimensional flat space time can thus be described starting from
\be
\vf(X) = \frac{1}{s!}\ { \vf}_{A_1 \ldots A_s}(X) \, {\tilde e}^{A_1}(X) \otimes \ldots
\otimes {\tilde e}^{A_s}(X) \, ,
\ee
and introducing the \emph{fixed and constant} auxiliary vector
$U^A=(u^a,v)$ one can again turn the attention to a function that is
homogeneous of degree $s$ in $U$,
\be
\vf(X,U^A) \equiv \vf(x,z,u^a,v) =  \frac{1}{s!}\ {\vf}_{A_1 \ldots A_s}(X) \ U^{A_1} \ldots U^{A_s}\,
.
\ee
Expanding in powers of $v$ then gives
\be \vf(X,U^A)=\sum_{r=0}^{s} {v^{r}\over r!}\ \vf_{s-r}(x,z,u^a) \,
.\ee

The symmetrized gradient of $\vf$ is a spin-$(s+1)$ field $\widehat{D\vf}$, that can also be turned into a function
of the auxiliary variable letting
\be (\widehat{D\vf})(X,U^A)={\widehat D} \vf(x,z,u^a,v) \, ,\ee
which in its turn defines $\widehat D$ as
\be
{\widehat D}=e^{\; -{z\over L}}\Big
[D+v\, \partial_z+{1\over L}
\left(u^2\, \partial_v-v \, u \cdot \partial_u \right)\Big] \, ,
\ee
with
\be
D = u^a\, e^\mu_a\, \left( \partial_\mu + \frac{1}{2}\ \omega_\mu{}^{ab} \, S_{ab} \right)\, .
\ee
The corresponding adjoint with respect to the scalar product (\ref{scalarflat}) in the $(d+1)$-dimensional flat
space time is then
\be
{\widehat D}^\dagger=e^{\; -\; {z\over L}}\left[ D^\dagger-\partial_v\partial_z-{1\over
L}\ \Big(d\, \partial_v+\partial_v \, u \cdot \partial_u-v\,
\partial_u \cdot \partial_u \Big)\right]\, ,
\ee
where the relation between $D$ and $D^\dagger$ was displayed in eq.~(\ref{ddag}).

Let us now consider a massless spin-$s$ field in the
$(d+1)$-dimensional flat space with the metric (\ref{metricz}). In
an arbitrary coordinate system,  its equation of motion in the
unconstrained formalism of \cite{fs,st,fms} reads
\be {\widehat {\cal F}}(X,U^A) - \frac{1}{2}\, {{\widehat D}^3\;
\alpha}(X,U^A) \, = \, 0 \ , \ee
where the flat Fronsdal operator in $(d+1)$ dimensions is
\be {\widehat {\cal F}}(X,U^A)= \left\{\left[{\widehat D},{\widehat
D}^\dagger\right]+ {\widehat D} \ {\widehat D}^\dagger+{1\over 2}\
({\widehat D})^2\,
\partial_U \cdot \partial_U \right\} \vf(X,U^A) \, .\label{frond+1}\ee
The resulting equations are invariant under the gauge
transformations
\be \delta \vf(X,U^A)={\widehat D}\Lambda(X,U^A),\quad
\delta\alpha(X,U^A)=
\partial_U \cdot \partial_U \ \Lambda(X,U^A)\, , \ee
that for the different components of the higher-spin field $\vf$ imply the relations
 \be \delta\vf_r(X)={e^{\; -{z\over
L}}}\left[D\Lambda_{r-1}(x)+ (s-r)\left(\partial_z-{r\over
L}\right)\Lambda_{r}(x) +{2\over L}\ g(x) \,
\Lambda_{r-2}(x)\right]\, . \label{phirgauge} \ee
When expressed in terms of $dS_d$ quantities, the flat-space
Fronsdal operator (\ref{frond+1}) becomes, after a lengthy but
straightforward calculation,
\ba {\widehat {\cal F}}(X,U^A)&=&e^{\; -{2z\over L}}\left\{ {\cal
F}+\left[\partial_z^2+{d-1\over L}\ \partial_z -{s\over
L^2}-{u^2\over 2}\ \left({s-4\over L^2}-{1\over L}\
\partial_z\right)\,
{\partial_u}\cdot {\partial_u}\right]\right. \vf \nonumber\\
&-&{v\over L}\ (s-2-L\, \partial_z)\ \left(  D^{\dagger}
+  D\,
{\partial_u}\cdot {\partial_u}\right)\vf \nonumber\\
&+&{v^2\over 2L^2}\ \left[(s-2)(s-3)+L^2\partial^2_z +(5-2s)L\partial_z\right]\
{\partial_u} \cdot  {\partial_u} \vf \nonumber\\
&+&{v\over L^2}\ \partial_v\left[-L^2\partial_z^2-(d-1)L\partial_z+
(d-1)u\cdot \partial_u+(u\cdot \partial_u)^2-u^2 \left({1\over
2}+ u\cdot \partial_u-L\partial_z\right)\partial_u\cdot \partial_u\right] \vf \nonumber\\
&+&{1\over L}\ \partial_v\left[
((2-d)-u\cdot \partial_u-L\partial z)  D+u^2(  D^\dagger
+  D\partial_u\cdot \partial_u)\right]\vf \nonumber\\
&+&{1\over 2L^2}\ \partial_v^2 \left[2u^2\left(
(1-d)-{1\over 2}\ L\partial_z-{3\over 2}\ u\cdot \partial_u+{1\over
2}\ u^2\partial_u\partial_u\right)+L^2  D^2\right]\vf
\nonumber\\
&+&{1\over 2L^2}\ v^2\partial_v^2\left[
 L^2\partial_z^2-L\partial_z+u\cdot \partial_u(1+u\cdot \partial_u)
 -2u\cdot \partial_uL\partial_z\right]\vf \nonumber\\
 &+&{v\over 2L}\ \partial_v^2\left[
 2L  D\partial_z-
   D(2+u\cdot \partial_u)\right]\vf \label{Fhat} \\
 &+&\left. \left[{v\over 2L^2}\ \partial^3_v
 u^2\, (2L\partial_z-2u\cdot \partial_u-3)
 +{1\over L}\ \partial_v^3\, u^2  D+{1\over 2L^2}\ \partial_v^4\, (u^2)^2 \right] \vf \right\}\,
 ,\nonumber
 \ea
where we have left implicit, for brevity, the argument $(X,U^A)$ of
the field $\vf$, and the Fronsdal operator for the given $dS$
background is
\be {\cal F}=\left( \Box_{dS}+  D  D^\dagger+{1\over 2}\ D^2
{\partial_u} \cdot {\partial_u}\right) \, \vf \, . \ee
Here $\Box_{dS}$ denotes the $dS$ d'Alembertian, and we used
eq.~(\ref{commut}), that for the $dS$ background becomes
\be [  D,  D^\dagger]=  D^a
D_a+{1\over L^2}\, \Big[s(s+d-2)-u^2 \ {\partial_u} \cdot
{\partial_u}\Big] \, .\ee
Taking into account the homogeneity of the composite field, one can
simplify to some extent these expressions making the two
replacements
\ba
&& v\; \partial_v \longrightarrow s-u \cdot \partial_u \, ,\nonumber \\
&& v^2\partial^2_v \longrightarrow (s-u \cdot \partial_u)(s-1-u \cdot \partial_u) \, .
\ea
\section{Free massive $dS$ fields}

One can now fix the gauge in such a way that $\vf(X,U^A)$ has no
$v$-dependent radial components,
\be
\partial_v
\vf(X,U^A)=0 \, ,
\ee
thus eliminating all internal Stueckelberg modes, as in Section 2.
The free wave equations then determine completely the compensator,
setting $\alpha=0$, and also imply the two conditions
\be
{\partial_u}\cdot {\partial_u}\ \vf(X,u^a)=0 \, , \qquad
D^\dagger\vf(X,u^a) = 0
\, .
\ee
Comparing the resulting expression with the expected form of the
massive spin-$s$ equation (see, \emph{e.g.}, eq.~(3.67) of
\cite{fms}, but with $L^2 \to - L^2$, since here we are discussing
the $dS$ case),
\be \Box_L \vf + \frac{2}{L^2}\ (s-1)(d+s-3) \ \vf - M^2 \vf = 0 \,
, \ee
where $\Box_L$ is the Lichnerowicz operator, defined for $AdS$ in
eq. (\ref{lichn}), one can read the resulting mass-shell condition,
\be \left[-M^2+{1\over L^2}\ (3-d-s)(2-s)\right] \vf(X,u^a) = \left(
\partial_z^2+{d-1\over L}\ \partial _z \right) \vf(X,u^a) \, . \ee
This simple differential equation determines the $z$-dependence  of
$\vf(X,u^a)$,
\be
\vf(X,u) \sim e^{{\mu\over L}z}\, ,\ee
with
\be
(ML)^2=(s-2-\mu)(d+s-3+\mu)\, .
\ee

The solutions of this mass-shell condition are then
\be \mu_{\pm}=-{d-1\over 2}\pm {i\over 2}\sqrt{(2ML)^2-(2s+d-5)^2}\,
\label{mass} . \ee
Notice that when $2ML<(2s+d-5)$ they are both real, and
\be
\mu_{\pm}=-{d-1\over 2}\pm {1\over 2}\sqrt{(2s+d-5)^2-(2ML)^2}\, .
\ee
Taking into account the measure $\sqrt{-g}\, g^{zz}=e^{{(d-1)z\over
L}}$, however, one can see that only the first solution, $\mu_+$, is
well behaved on the light cone, $z=-\infty$ or $R=0$, of the
original $(d+1)$-dimensional flat space, and is thus acceptable,
while the other, $\mu_-$, should be rejected. On the other hand,
when $2ML\geq(2s+d-5)$ the two solutions are related to one another
by complex conjugation, and therefore are both acceptable. This
pattern reflects the more familiar one found in the more
conventional toroidal reduction. Even in that case a single periodic
wave function, the constant, determines the massless mode, while
pairs of periodic wave functions, characterized by opposite momenta,
determine the massive ones.

\subsection{Discrete states and ``partially massless'' $dS$ fields}

As we anticipated, for generic values of $\mu$, fixing the
Stueckelberg gauge symmetries one can set $\vf_r$=0 for
$r=0,1,\dots, s-1$, and this fixes the gauge completely. Strictly
speaking, however, this procedure is not quite possible when
$\mu+1=k$, with $k=s-1, s-2,\dots ,0$. In these cases a residual
gauge invariance emerges \cite{partlymass}, that is associated with
spin-$k$ gauge parameters $\Lambda_k$, together with a consequent
reshuffling of the propagating degrees of freedom. This can be seen
extracting the $z$ dependence while taking into account the overall
factor in eq.~(\ref{phirgauge}),
\be
\Lambda_r \sim e^{(\mu+1) \frac{z}{L}} \qquad \forall r \, ,
\ee
so that the gauge transformations can be simplified and become
 \be \delta\vf_r={e^{\;-\; {z\over
L}}}\left[D\Lambda_{r-1}+ (s-r)\left({{\mu+1 - r}\over
L}\right)\Lambda_{r} +{1\over 2 L}\ g \, \Lambda_{r-2}\right]\, ,
\qquad (r=0, \ldots, s) \, . \label{phirgauge2} \ee

Notice that, if
\be \mu = k-1 \,, \ee
the middle coefficient in eq.~(\ref{phirgauge2}) vanishes for $r=k$,
so that the corresponding parameter, $\Lambda_{k}$, has no effect on
$\delta \vf_k$. This allows to determine, recursively, corresponding
values for the other parameters $\Lambda_{l}$, for $l >k$, capable
of keeping all the $\vf_r$ for $r \neq k$ fixed at their vanishing
values, for $r=0,\ldots,s-1$. The resulting residual gauge
transformation of $\vf_s$ finally reads
\be
\delta \vf_s = {e^{\;-\; {z\over L}}}\, \left[   D\Lambda_{s-1}+
{1\over L}\ u^2\, \Lambda_{s-2}\right]\, ,
\ee
where $\Lambda_{s-1}$ and $\Lambda_{s-2}$ are determined from $\Lambda_k$ according to
\be
\Lambda_{k+1}={L\over s-k-1}\    D\Lambda_{k}\, ,
\ee
and
\be
\Lambda_{t}={L\over(t-k)(s-t)}\ \left[   D\Lambda_{t-1}
+{1\over L}\ u^2\, \Lambda_{t-2}\right]\, , \qquad
(t=k+2,k+3,\dots,s-1) \, .
\ee
The corresponding masses are
\be
 (ML)^2=(s-1-r)(d+s+r-4)\, , \quad (r=s-1,\dots,0)\, ,
\ee
and in particular $r=s-1$ corresponds to the massless case, while
the full sequence of values of $(ML)^2$ for the discrete states is
$0, d+2s-6, 2(d+2s-7), 3(d+2s-8), \dots, (s-1)(d-4+s)$.

Let us stress that, from the $(d+1)$-dimensional vantage point, what
happens is a mere redistribution of degrees of freedom: for the
special masses above what is usually a Stueckelberg field becomes a
propagating field, which takes up precisely the degrees of freedom
lost by the ``partially massless" field $\vf$. Nothing special
happens, in fact, in the $(d+1)$-dimensional flat space, for these
special values of the mass. This is to be contrasted with the
conventional way of looking at partial masslessness directly in $d$
dimensions, where one starts with a massive field and discovers that
a residual gauge symmetry emerges for special values of $(ML)^2$.

In four dimensions, the discrete values for the mass $M$ that we
have thus identified correspond to the unitary representations
$\pi_{p,q}$, with $1 \leq q \leq  p$, determined by Dixmier in
\cite{rep1}, with the choices $p=s$ and $q=k+1$. In order to compare
with the results of that paper, let us note that the second-order
Casimir operator $I_2$ of the de Sitter group is related to the
masses here defined by
\be
I_2+2(s^2-1)-(ML)^2=0 \, .
\ee

The additional representations called $\pi_{p,0}$ by Dixmier
\cite{rep1} are special scalar representations, minimally massless
for $p=0$ and tachyonic for the other values. The other bosonic
unitary representations in $\cite{rep1}$ are characterized by an
integer $p$, the spin $s$ and a continuous positive label $\sigma$,
and are called $\nu_{p,\sigma}$: their masses are given by
\be
(ML)^2=(s^2)-s+\sigma \, .
\ee
Notice that these values are bounded from below by the masses of the
discrete states, the last of which is also determined in four
dimensions by this expression for $\sigma=0$. This last type of
representation corresponds to the unitary massive fields. The
representations of the higher-dimensional de Sitter groups are more
complicated, since they also involve mixed-symmetry fields, and were
studied in\cite{rep2}. The construction of $dS$ quantum field
theories starting from unitary irreducible representations of
$SO(1,D)$ was recently considered in \cite{Joung:2006gj}.

\section{Coupling to a $d$-dimensional conserved current}

We can now turn to the central topic of this paper, the coupling of
massive higher-spin fields to external currents in a de Sitter
background.

\subsection{Conserved $dS$ currents from flat $(d+1)$-dimensional currents}
The  $(d+1)$-dimensional external current $J$ couples to a massless
field of the unconstrained formulation of \cite{fs,st,fms}, and
therefore should be conserved, a condition that in the present
notation reads
\be
{\widehat D}^\dagger {\widehat J}(X,U^A)=0\, ,
\ee
and in $d$-dimensional terms becomes
\be \left[
D^\dagger-\partial_v\partial_z-{1\over
L}\left(d\partial_v+\partial_v u \cdot {\partial_u}-v {\partial_u}
\cdot {\partial_u}\right) \right]\, J(X,U^A)=0\, . \label{diffcurr} \ee
Notice that, with the chosen exponential $z$-dependence, that we are
factoring out of all expressions,
\be
\partial_z \ \to \ \frac{1}{L}\ (\mu-2) \ee
when acting on $J$, where the ``shift" is due to the pre-factor in
eq.~(\ref{Fhat}). The $v$-dependence of the current can be made
explicit integrating eq.~(\ref{diffcurr}), and is fully encoded in
the expression
\be J(x,U^A)=e^{\left(\mu+d+u  \cdot \partial_u-2\right)^{-1}\left(Lv
D^\dagger +{v^2\over2}\partial_u \cdot \partial_u\right)}\
J_s(x,u^a)\, . \label{consd+1}\ee

If one adds the further condition that the $d$-dimensional current
be also conserved, so that
\be
D^\dagger J_s(x,u^a)=0 \,,
\ee
eq.~(\ref{consd+1}) reduces to
\be
J(x,U^A)=\left[1+ \sum_{n=1}^{[\frac{s}{2}]}{(v^2\partial_u \cdot
\partial_u)^n\over 2^n\; n!\,  (\mu
+d+s-4)(\mu+d+s-6)\dots(\mu+d+s-2n-2)} \right]J_s\, , \ee
and letting
\be \nu=\frac{1}{2}\, (\mu+d+s) \, ,\ee
it can be recast in the more convenient form
\be J(x,U^A)= \sum_{n=0}^{[\frac{s}{2}]}\, (-1)^n {1\over 2^{2n}\;
n!(2-\nu)_n}\ (v^2\partial_u \cdot
\partial_u)^nJ_s(x,u^a)\, . \ee
The $v$-dependence of this current reflects the fact that its radial
components do not vanish identically, and is an important novelty with respect to the flat
Kaluza-Klein reduction reviewed in Section 2.

The  $m$-th trace of the current $J$ in $(d+1)$ dimensions will also
be useful later. It can be derived from the previous expression by
manipulations similar to those illustrated in Appendix B, or by an
inductive argument, and is given by
\be
(\partial_U \cdot \partial_U)^mJ(x,U^A)=\left({3\over
2}-\nu\right)_{m}
 \sum_{n=0}^{\left[\frac{s}{2}\right] -m }\!\! {{(-1)^n} \over {2^{2n}\; n!(2-\nu)_{n+m}}}\
(v^2\partial_u \cdot \partial_u)^n(\partial_u \cdot
\partial_u)^mJ_s(x,u^a)\, .
\ee

\subsection{The field equations}

In terms of the gauge-invariant extension of the Fronsdal operator,
\be {\cal A}(x,U^A)={\widehat {\cal F}}(x,U^A)-\frac{1}{2}\ {\widehat
D}^3\alpha(x,U^A) \, ,\label{acurv} \ee
proceeding as in \cite{fms} the field equation with an external
current in the $(d+1)$-dimensional Minkowski space can be turned
into
\be
{\cal A}(x,U^A)=\sum_{n=0}^{[s/2]}{\rho_n(d-1,s)\over 2^n n!}\
(U^2)^n (\partial_U \cdot \partial_U)^n J\equiv f(x,U^A)\, ,
\label{eqauj} \ee
where
\be
 (\partial_U \cdot \partial_U)^2 {\cal A}(x,U^A)=0\, , \ee
since ${\cal  A}$ is doubly-traceless in $(d+1)$ dimensions, after
using eq.~(\ref{nodtr}). Here $J$ is the $(d+1)$-dimensional
conserved current, and as we have seen
\be \rho_n(d-1,s)={1\over 2^n\left({5\over 2}-\zeta\right)_{n}}\ ,\ee
where $\zeta$ is defined in eq.~(\ref{zeta}).

Substituting for $J$ its expression in terms of the de Sitter
current $J_s(x,u^a)$ then gives
\be
f(x,U^A)=\sum_{n=0}^{[\frac{s}{2}]}{(u^2+v^2)^{n}\over{2^{2n}\;
n!}\left({5\over 2}-\zeta\right)_{n}}\, \left({3\over
2}-\nu\right)_{n}
 \sum_{p=0}^{[\frac{s}{2}]-n}{(-1)^p \over 2^{2p}\; p!\,
 (2-\nu)_{p+n}}\
(v^2)^p\, (\partial_u \cdot
\partial_u)^{n+p} J_s \, .
\label{fjs} \ee

After rearranging this expression as in Appendix B, one can recast
eq.~(\ref{fjs}) in the rather compact form
\be f(x,U^A)\, =\, \sum_{r=0}^{[\frac{s}{2}]} \ \sum_{m=0}^{[\frac{s}{2}]-r} b(r,m){(v^2)^r\over
(2r)!}\ {(u^2)^m}\ (\partial_u \cdot \partial_u)^{r+m}J_s(x,u^a)\,
,\label{fff} \ee
where the coefficients $b(r,m)$ are
\be
b(r,m)=(-1)^r{(2r)!\over 2^{2(r+m)}r!\, m!}\
{\left({3\over 2}-\nu\right)_m\left(\nu-\zeta+1\right)_r\over
\left(2-\nu\right)_{r+m}\left({5\over 2}-\zeta\right)_{r+m}}\ .\label{bb}
\ee

\subsection{Determining the compensator $\alpha$}

As in the flat case, let us work in the gauge $\partial_v\vf=0$,
with no radial components for the spin-$s$ gauge field. Our goal is
to determine the current-exchange amplitude, and to this end we
should first determine the compensator $\alpha$, the trace and the
divergence of $\vf$, up to $d$-dimensional gradients, in terms of
the external de Sitter current. The $v$-independent part of the
field equation then determines the effective kinetic operator
relevant for the current exchanges,
\be {\cal K}(x,u^a) \equiv \left( \Box_L + \frac{2}{L^2}\
(s-1)(d+s-3) - M^2 \right) \vf(x,u^a) \, . \label{defk}\ee
In this subsection we begin by expressing
\be
\alpha(x,U^A)=\sum_{i=0}^{s-3} \alpha_i(x,u^a)\, \frac{v^i}{i!}
\ee
in terms of $J_s(x,u^a)$, and we shall see shortly that $\alpha$
contains only \emph{odd} powers of $v$. In fact, we shall first
express the components $\alpha_i$ in terms of those of
\be f(x,U^A)=\sum_{r=0}^{\left[\frac{s}{2} \right]}
f_{2r}(x,u^a) \, \frac{v^{2r}}{(2r)!} \ , \ee
that we already related to $J_s(x,u^a)$ in eq.~(\ref{fff}), so that
 \be f_{2r}(x,u^a)=
\sum_{m=0}^{\left[\frac{s}{2}\right]-r} b(r,m)\, (u^2)^m(\partial_u \cdot \partial_u)^{r+m}J_s(x,u^a)\, ,
\label{fn}\ee
with $b(r,m)$ given in eq. (\ref{bb}).

As can be seen from eqs.~(\ref{Fhat}) and (\ref{eqauj}), ${\cal F}$
does not involve any terms of order $v^3$ or higher, and therefore
the field equation (\ref{eqauj}) implies the conditions
\be
\partial^3_v\left(\frac{1}{2}\ {\widehat D}^3\alpha +f(x,U^A)\right)=0\,
.
\ee
We shall see shortly that this determines completely ${\widehat
D}^3\alpha$, and consequently, as we anticipated, $\alpha$ will only
contain odd powers of $v$. This is due to the fact that $f(x,U^A)$
is even in $v$, which will then be the case also for ${\widehat
D}^3\alpha$ that, up to $d$ dimensional gradients, is given by
\ba {\widehat D}^3\alpha(X,U^A)&=&e^{\;  {(\mu - 2) z \over L}}
\left\{{v^3\over
L^3}\, (\mu-1-u \cdot \partial_u)(\mu-u\cdot \partial_u)(\mu+1-u\cdot \partial_u)\; \right. \nonumber\\
&+& \left. {3v\over L^3}\, u^2\left[(s-3-u\cdot \partial_u)
(\mu-1-u\cdot \partial_u)^2
+(\mu+1-u\cdot \partial_u)(\mu-1-u\cdot \partial_u)\right] \right.\nonumber\\
&+&\left. {3\partial_v\over L^3}\, (u^2)^2\left[(s-3-u\cdot
\partial_u)(\mu-2-u\cdot \partial_u) +2\right]+{\partial^3_v\over
L^3}\, (u^2)^3 \right\}\alpha(x,U^A)\, , \label{d3alpha} \ea
where different arguments, $x$ and $X$, are present on the two sides
of this equation since we have made the dependence on $z$ fully
explicit. In deriving this and the following expressions we are
using repeatedly the further condition
\be
\Big(v \partial_v + u \cdot \partial_u\Big)\; \alpha(x,U^A) = (s-3) \; \alpha(x,U^A) \, ,
\label{homa}
\ee
which reflects the degree of homogeneity, and thus the spin, of the
compensator $\alpha_{A_1 \ldots A_{s-3}}$. As a result, letting
\be
g=-2\, L^3\, \partial_v^{\;3}\, f(x,U^A)\, ,\label{gdef}
\ee
the equation for $\alpha(x,U^A)$ can finally be written in the form
\be
\Big[A_0+A_2\; \partial_v^{\;2}+A_4\; \partial_v^{\;4}+A_6\; \partial^{\;6}_v\Big]\; \alpha(x,U^A)=g\, ,
\ee
with
\ba
A_0&=&(s-u\cdot \partial_u)(s-1-u\cdot \partial_u)(s-2-u\cdot \partial_u)(\mu-1-u\cdot \partial_u)
(\mu-u\cdot \partial_u)(\mu+1-u\cdot \partial_u)\,  ,\nonumber\\
A_2&=&3u^2(s-2-u\cdot \partial_u)\Big[(s-3-u\cdot \partial_u)
(\mu-1-u\cdot \partial_u)^2
+(\mu+1-u\cdot \partial_u)(\mu-1-u\cdot \partial_u)\Big]\, , \nonumber\\
A_4&=&3(u^2)^2\Big[(s-3-u\cdot \partial_u)(\mu-2-u\cdot \partial_u)
+2\Big]\, , \nonumber\\A_6&=&(u^2)^3\, .
\ea

In order to determine $\alpha$ explicitly, let us take into account
eq.~(\ref{homa}) so as to replace $u \cdot \partial_u$ with $v
\partial_v$, and let us define
\be
a = \mu - s \equiv 2(\nu - \zeta) \,.
\ee
Eq. (\ref{d3alpha}) can then be fully expressed in terms of the four
quantities
\ba
E_0(t)&=&(1+t)(2+t)(3+t)(a+2+t)(a+3+t)(a+4+t)\, ,\nonumber\\
E_2(t)&=&3(1+t)[t(a+2+t)^2+(a+4+t)(a+2+t)]\, ,\nonumber\\
E_4(t)&=&3t(a+1+t)+6\, , \nonumber \\
E_6(t)&=&1 \, ,
\ea
and let us also define
\be \tilde
E_2(t)=-\ {E_2(t)\over E_0(t-2)}\, ,\quad \tilde E_4(t)=\ -\
{E_4(t)\over E_0(t-4)}\, ,\quad \tilde E_6(t)=\ -\ {1\over
E_0(t-6)}\, , \ee
and
\be
\gamma={1\over E_0(v\partial_v)}\ g\, . \label{gammag} \ee
Here $g$ is given in eq.~(\ref{gdef}), and $t$ will be shortly
identified with $v \partial_v$. The equation for $\alpha$ thus reads
\be \left[1-\sum_{i=1}^{3}(u^2)^i\partial_v^{2i}\tilde
E_{2i}(v\partial_v)\right]\alpha(x,U^A)= \gamma\, ,
\label{alphagamma} \ee
where we have made use of the relation
\be v\partial_v
\left(\partial_v\right)^i=\left(\partial_v\right)^i(v\partial_v-i)\, . \ee

One can now invert eq.~(\ref{alphagamma}) and write the solution in the form
\ba \alpha(x,U^A)&=& \sum_{n\geq 0}\ \left[\sum_{i=1}^{3}(u^2)^i\partial_v^{2i}
\tilde E_{2i}(v\partial_v)\right]^n\gamma \\
&=&\sum_{n\geq 0}\sum_{i_1,\dots,i_n=1}^{3}
(u^2\partial_v^2)^{i_1+i_2+\dots+i_n}\ \tilde
E_{2i_1}(v\partial_v-2i_2-\dots-2i_n)
\dots \tilde E_{2i_n}(v\partial_v) \ \gamma\, .\nonumber
\ea
This result embodies explicit expressions for all the $\alpha_i$,
\ba
\alpha_i&=&\gamma_i+\sum_{n\geq 1}(u^2)^{n}\
\sum_{p\geq 1}\sum_{i_1+\dots +i_p=n}
\tilde E_{2i_1}(i+2i_1)\tilde E_{2i_2}(i+2i_1+2i_2)\dots
\dots \tilde E_{2i_p}(i+2n)
\gamma_{i+2n}\label{alp}\\
&=&\gamma_i+\sum_{n\geq 1}(u^2)^{n}\ \sum_{p\geq 1}\sum_{i_1+\dots+
i_p=n}(-1)^p { E_{2i_1}(i+2i_1) E_{2i_2}(i+2i_1+2i_2)\dots
E_{2i_p}(i+2n)\over E_0(i)E_0(i+2i_1)\dots
E_0(i+2i_1+\dots+2i_{n-1})}\ \gamma_{i+2n} \,  , \nonumber \ea
where, from eq.~(\ref{gammag}),
\be \gamma_i=-2L^3\frac{f_{i+3}}{E_{0}(i)} \, .\ee

Having determined $\alpha(x,U^a)$ from the previous equation, one
can now obtain the zeroth-order and second-order contributions to
${\widehat D}^3\alpha$, that enter the $\vf$ equations at the same
order in $v$, making use of eq.~(\ref{d3alpha}), and the end result
is
\ba
\!\!\!\!\!\!\!\!\!L^3\, {\widehat D}^3\alpha\!\!&=&\!\!3(u^2)^2(a+4)\alpha_1+(u^2)^3\alpha_3\nonumber\\
&+&\!\!\! \left[12u^2(a+3)(a+4)\alpha_1+3(u^2)^2(3a+14)\alpha_3
+(u^2)^3\alpha_5\right]{v^2\over 2}+ {\cal O}(v^4) \, .
\ea

\section{The $d$-dimensional equation}

We are now ready to analyze the three remaining equations, the
$(d+1)$-dimensional equations at order $v^k$ with $k \leq 2$. The
final aim will be to relate ${\cal K}(x,u^a)$ of eq.~(\ref{defk}) to
the external current $J_s(x,u^a)$, up to $d$-dimensional gradients.
We shall proceed in steps, obtaining first the expression in terms
of $f(x,U^A)$, in eq.~(\ref{prop1}), and then, in the next section,
the explicit dependence on $J_s(x,u^a)$.

The order-$v^2$ terms in the equation of motion determine the trace
of $\vf$, so that, making use of eq. (\ref{Fhat}),
\be \frac{(a+2)(a+3)}{L^2}\
{\partial_u\cdot \partial_u}\vf(x,u^a)=f_2 +{u^2 \over 2L^3}
\Big[12(a+3)(a+4)\alpha_1+3(u^2)(3a+14)\alpha_3
+(u^2)^2\alpha_5\Big]\, , \label{traced} \ee
while the terms of order $v$ in the field equation lead to the condition
\be \left(  D^{\dagger} +  D\;
{\partial_u \cdot \partial_u}\right) \vf(x,u^a)=0\, , \ee
that determines the divergence of $\vf$.

Finally, the order-$v^0$ part of the field equation, after using the
previous results, is the effective $d$-dimensional equation we are
after, that up to pure gradients, which do not contribute to the
current exchange, reads
\be {\cal K}(x,u^a)+ \left[ {u^2\over
2L^2}\, (a+2)\; \partial_u\cdot \partial_u\right]\vf(x,u^a)
=f_0+{3\over 2L^3}\ (u^2)^2(a+4)\alpha_1+{(u^2)^3\over {2L^3}}\
\alpha_3\, , \ee
where ${\cal K}$ is defined in eq.~(\ref{defk}). Replacing the trace
$\partial_u
\cdot
\partial_u\vf $ by its explicit form obtained from
eq.~(\ref{traced}) gives
\be {\cal K}(x,u^a) = f_0-u^2{f_2\over 2(a+3)}-{3\over
2L^3}\ (u^2)^2(a+4)\alpha_1-
{(u^2)^3(7a+36)\alpha_3+(u^2)^4\alpha_5\over 4(a+3)L^3}\, . \ee
Notice that the $\alpha_i$ vanish for $s< 3+i$, so that all the
three contribute only for $s\geq 8$. Replacing the $\alpha_i$ with
their expressions (\ref{alp}) in terms of the $f_i$ yields
\ba {\cal
K}(x,u^a)&=& f_0-u^2{f_2\over 2(a+3)} +3(u^2)^2(a+4)\left[
{f_4\over E_0(1)}+{\cal K}_0\right]\nonumber\\
&+&(u^2)^3{(7a+36)\over 2 (a+3)}\left[
{f_6\over E_0(3)}+{\cal K}_1\right]
+(u^2)^4{1\over 2(a+3)}\left[
{f_8\over E_0(5)}+{\cal K}_2\right]\, , \label{kappa}
\ea
with
\be
{\cal K}_j=\!\!\!\!\!\!\!\!\! \sum_{n\geq 1,p\geq 1,i_1+\dots i_p=n}\!\!\!\!\!\!\!\!\!\!\!\!\!\!\! (u^2)^{n}
{(-1)^p E_{2i_1}(1+2j+2i_1) E_{2i_2}(1+2j+2i_1+2i_2)...
 E_{2i_p}(1+2j+2n)\over E_0(1+2j)E_0(1+2j+2i_1)E_0(1+2j+2i_1+2i_2)...
E_0(1+2j+2n)}f_{4+2j+2n}\, .\nonumber
\ee

This expression looks indeed rather clumsy. Remarkably, however, it
can be largely simplified, since the sums over all the $i_i$ can be
performed explicitly. The final result, proved in Appendix C, is
simply
\be {\cal K}(x,u^a)=\sum_{r=0}^{\left[ \frac{s}{2} \right]}\
(u^2)^r\, \lambda_r \, f_{2r}\, ,\label{prop1} \ee
where the coefficients are
\be
\lambda_r=(-1)^r\, {1\over 2^{2r} r!\left(\frac{a+3}{2}\right)_r} = (-1)^r\, {1\over 2^{2r} r!\left(\nu-\zeta+{3\over
2}\right)_r} \, .\label{lam} \ee
This is the main result of this Section.

Before considering the general case, let us now examine the massless and flat limits of the
propagator. First, the massless limit is attained for $\mu\rightarrow s-2$, which implies that
$\zeta-\nu\rightarrow 1$. From eq.~(\ref{bb}), we see that in this limit all the $b(r,m)$ with $r \neq0$ vanish, while
\be \lim_{\mu \to s-2} \ b(0,m) =  {1\over
2^{2m}m!\, (3-\zeta)_m}\, , \ee
so that
\be \lim_{\mu \to s-2} {\cal K}(x,u^a) = \sum_{m=0}^{\left[\frac{s}{2} \right]} {1\over
2^{2m}m!\, (3-\zeta)_m}\ (u^2)^m(\partial_u \cdot
\partial_u)^mJ_s(x,u^a)\, , \ee
which is precisely the massless propagator of \cite{fms}. In other
words, \emph{the vDVZ discontinuity is absent for all $s$}, exactly
as was found to be the case in \cite{porrati} for $s=2$. Notice
that, strictly speaking, this limit should be taken after a proper
continuation to $AdS$ not to leave the region of unitarity.

On the other hand, the flat limit is obtained for $L\rightarrow
\infty$ with finite $M$, which implies that $\nu\rightarrow
i\infty$. From eq.~(\ref{bb}) one can deduce the limiting behavior
of the coefficients $b(r,m)$,
\be
\lim_{\nu\to i \infty}  b(r,m) = {(2r)!\over2^{2(r+m)}\, r!\, m!\, \left({5\over
 2}-\zeta\right)_{r+m}}\, ,\ee
while eq.~(\ref{lam}) shows that all the $\lambda_r$ coefficients
tend to zero for $r \neq  0$. One can then readily deduce the
limiting behavior of the propagator
\be \lim_{\nu \to i \infty} {\cal K}(x,u^a)\rightarrow \sum_{m=0}^{\left[\frac{s}{2} \right]} {1\over
2^{2m}m!} \left({5\over 2}-\zeta\right)_{m}\
(u^2)^m(\partial_u.\partial_u)^mJ_s(x,u^a)\, ,\ee
which is precisely the massless $(d+1)$-dimensional propagator in
Minkowski space that one can also derive letting $M \to 0$ in
eq.~(\ref{fmsfinm}).

\section{The $(A)dS$ current exchanges}

We can finally express the coefficients $f_{2r}$ in
eq.~(\ref{prop1}) in terms of the $d$-dimensional current
$J_s(x,u^a)$ and obtain the $dS$ current-exchange amplitudes for all
spin-$s$ bosonic fields.

\subsection{Some useful identities}

The expressions that were obtained so far, and in particular
eq.~(\ref{prop1}), depend on $\nu$, that in its turn determines the
mass via eq.~(\ref{mass}) and we rewrite for convenience in the form
\be \nu={2\zeta+1\over 4}\pm \frac{i}{2}\, \sqrt{(ML)^2-\left(\zeta-{5\over
2}\right)^2}\, . \label{masshell}\ee
where $\zeta$ is defined in eq.~(\ref{zeta}). As a result,  both the
higher dimensional current $J$ and the compensator field $\alpha$
\emph{are not analytic} in $(ML)^2$. On the other hand, as we shall
see, the propagator ${\cal K}$ is always a rational function of
$(ML)^2$, as was shown to be the case for $s=2$ in
\cite{higuchi,porrati}. Proving this result, however, requires some
intermediate steps, to which we now turn.

Let us begin by displaying a few expressions that will prove useful
in exhibiting the mass dependence of the propagator. To this end,
let us recall that the mass shell condition reads
\be (ML)^2=2(2\nu-3)(\zeta
-\nu-1)=-4\nu^2+2\nu(2\zeta+1)-6(\zeta-1)\, , \label{masshell2}\ee
and implies that
\be 4\left({3\over
2}-\nu+c\right)(\nu-\zeta+1+c)=(ML)^2+2c(2c+5-2\zeta)\, , \ee
for any $c$. One can therefore deduce the useful relations
\be  \left({3\over
2}-\nu\right)_n(\nu-\zeta+1)_n={1\over 2^{2n}}\ \prod_{j=0}^{n-1}
\Big[(ML)^2+2j(2j+5-2\zeta)\Big]\equiv h_n(\nu,\zeta)\, , \label{hn} \ee
\be
\left(2-\nu\right)_n\left(\nu-\zeta+{3\over 2}\right)_n={1\over
2^{2n}}\ \prod_{j=0}^{n-1}
\Big[(ML)^2+2(2j+{1})(j+3-\zeta)\Big]\equiv g_n(\nu,\zeta) \, , \label{gn}
\ee
where $h_n$ and $g_n$ are manifestly \emph{polynomials} in $(ML)^2$
and $\zeta$. This result can also be understood from a different
vantage point. To this end, let us define the new variables
\be \label{gd}
\begin{split}
&\gamma \, = \, \frac{1}{2} \, \left(\frac{5}{2} \, - \, \zeta\right) \, , \\
&\delta \, = \, \pm{\frac{i}{2}}\, \sqrt{(ML)^2 \, - \, \left({5\over
2} \, - \, \zeta \right)^2}\, ,
\end{split}
\ee
so that the arguments of the Pochhammer symbols in eqs.~(\ref{hn})
and (\ref{gn}) can be conveniently expressed as
\be
\begin{split}
&{3\over2}\, -\, \nu \, = \,  \gamma \, - \, \delta \,
,\qquad\quad\quad
\nu \, - \, \zeta \, +\, 1 \, = \, \gamma \, + \, \delta \, ,\\
&2 \, - \, \nu \, = \, \gamma \, - \, \delta \, + \, \frac{1}{2} \,
,\qquad \nu \, - \, \zeta \, + \, {3\over 2} \, = \, \gamma \, + \,
\delta + \, \frac{1}{2}\, .
\end{split}
\ee
This makes it possible to rewrite $h_n$ and $g_n$ in the form
\be
\begin{split}
& h_n \, = \, (\gamma \, - \, \delta)_n \, (\gamma \, + \, \delta)_n \, , \\
& g_n \,  = \,
\left[(\gamma + \frac{1}{2}) \, - \, \delta\right]_n \, \left[(\gamma + \frac{1}{2})\,
+ \, \delta\right]_n .
\end{split}
\ee
Eqs. (\ref{hn}) and (\ref{gn}) then follow as a consequence of a
generalisation of the standard formula for the difference of two
squares, \be (\gamma \, - \, \delta)_n \, (\gamma \, + \, \delta)_n
\, = \, \prod_{i = 0}^{n -1} \, \{(\gamma + i)^2\, - \, \delta^2) \,
, \ee where the explicit dependence only on even powers of $\delta$
is again manifest. Let us stress that the roots of $h_n$ are the
masses of the \emph{odd discrete states}, while the roots of $g_n$
are the masses of the \emph{even discrete states}.

\subsection{Expansion of the propagator: the first few cases}

From eqs.~(\ref{prop1}), (\ref{lam}) and (\ref{fn}) one can see that the propagator
${\cal K}$ takes form
\be {\cal
K}(x,u^a)=\sum_{n=0}^{\left[\frac{s}{2}\right]}k_{n}{\left(u^2\right)^n}
(\partial_u.\partial_u)^nJ_s(x,u^a)\, , \ee
with
\be
k_n=\sum_{r=0}^{n}(-1)^r\,b(r,n-r)\, {1\over 2^{2r
}r!\left(\nu-\zeta+{3\over 2}\right)_r}\, . \ee
Using the expression for the coefficients $b(r,m)$ in (\ref{bb}), one
can obtain
\be
k_n={1\over 2^{2n} \left(2-\nu\right)_{n}\left({5\over
2}-\zeta\right)_{n}}\, \sum_{r=0}^{n}{(2r)!\over
2^{2r}(r!)^2(n-r)!} \ {\left({3\over
2}-\nu\right)_{n-r}\left(\nu-\zeta+1\right)_r\over
\left(\nu-\zeta+{3\over 2}\right)_r}\ .\label{sn}\ee
We shall examine shortly the general expression for the coefficients
$k_n$. As we shall see, they are all rational fractions of $(ML)^2$,
although this is clearly not evident at first sight, but let us
begin by examining the first four terms. The first two are
\be k_0=1,\quad k_1={4h_1+5-2\zeta\over 8 g_1\left(5-2
\zeta\right)}\, , \ee
where $h_1$ and $g_1$ were defined respectively in (\ref{hn}) and
(\ref{gn}). As we have seen in the previous subsection, $h_1$ is
linear in $(ML)^2$, so that
\be k_1=- {1\over 2(2\zeta-5)}\
{(ML)^2-2\zeta+5\over(ML)^2-2\zeta+6}\ \label{k1}\ee
is indeed a rational function of $(ML)^2$. In a similar fashion,
\be k_2={1\over 2^{4} \left(2-\nu\right)_{2}\left({5\over
2}-\zeta\right)_{2}} \left[{1\over 2}\left({3\over
2}-\nu\right)_{2} +{1\over 2}\left({3\over
2}-\nu\right){\left(\nu-\zeta+1\right)\over
\left(\nu-\zeta+{3\over 2}\right)}+{3\over
8}{\left(\nu-\zeta+1\right)_2\over \left(\nu-\zeta+{3\over
2}\right)_2}\right] \, , \ee
but after some rearrangements, using also eqs.~(\ref{hn}) and
(\ref{gn}), one can turn this expression into
\be
k_2={(ML)^4-4(ML)^2(2\zeta-7)+3(2\zeta-5)(2\zeta-7)
\over 8(2\zeta-5)(2\zeta-7)[(ML)^2-2\zeta+6]
[(ML)^2-6\zeta+24]} \ ,\ee
that is again a rational function of $(ML)^2$.

The next case is more involved, and therefore let us examine it in
detail, since
\ba
k_3=b(0,3)&-&{b(1,2)\over 2(2\nu-2\zeta+3)} +{b(2,1)\over 8
(2\nu-2\zeta+3)(2\nu-2\zeta+5)} \nonumber \\
&-&{b(3,0)\over 48\;(2\nu-2\zeta+3)(2\nu-2\zeta+5) (2\nu-2\zeta+7)}\, . \ea
We would like to show that this expression is also a function of $(ML)^2$ and
$\zeta$. After replacing the $b(r,m)$ by their expressions,
$k_3$ can be written in the form
\be k_3={{\cal N}_3\over {\cal D}_3}\  , \ee
with
\be {\cal D}_3=2^{9}3!\left({5\over
2}-\zeta\right)_3(2-\nu)_3\left(\nu-\zeta+{3\over 2}\right)_3
=2^{9}3!\left({5\over 2}-\zeta\right)_3g_3\, , \ee
and
\ba
{\cal N}_3&=&8\left({3\over 2}-\nu\right)_3\left(\nu-\zeta+{3\over
2}\right)_3 +12\left({3\over
2}-\nu\right)_2\left(\nu-\zeta+{5\over 2}\right)_2
\left(\nu-\zeta+1\right)\nonumber\\
&+&18\left({3\over 2}-\nu\right)\left(\nu-\zeta+{7\over 2}\right)
\left(\nu-\zeta+1\right)_2
+15\left(\nu-\zeta+1\right)_3\,  .
\ea
One can now use the binomial identity (\ref{new}) for Pochhammer
symbols to obtain
\ba
&& \left(\nu-\zeta+{3\over 2}\right)_3=\sum_{k=0}^3{6\over k!(3-k
)!}
\left({1\over 2}\right)_k
\left(\nu-\zeta+1\right)_{3-k} \, ,\nonumber \\
&& \left(\nu-\zeta+{5\over 2}\right)_2
=\sum_{k=0}^2{2\over k!(2-k
)!}
\left({1\over 2}\right)_k
\left(\nu-\zeta+2\right)_{2-k}\, ,
\ea
so that the numerator becomes
\ba
{\cal N}_3&=&8h_3+12\left({7\over 2}-\nu\right)h_2+18\left({5\over 2}-\nu\right)_2
h_1+15\left({3\over 2}-\nu\right)_3
+12\left(\nu-\zeta+3\right)h_2 \\
&+&12h_2+9\left({5\over 2}-\nu\right)h_1
+18\left(\nu-\zeta+2\right)_2h_1
+9\left(\nu-\zeta+2\right)h_1+15\left(\nu-\zeta+1\right)_3\, .
\nonumber\ea
The next step is to use eq.~(\ref{new}) to obtain
\ba
\left({5\over 2}-\nu\right)_2+\left(\nu-\zeta+2\right)_2&=&
\left({9\over 2}-\zeta\right)_2-2\, {h_2\over h_1}\\
\left({3\over 2}-\nu\right)_3+
\left(\nu-\zeta+1\right)_3&=&\left({5\over 2}-\zeta\right)_3
-3\left({9\over 2}-\zeta\right)h_1\, ,
\ea
so that the numerator finally becomes
\be
{\cal N}_3=8h_3+12\left({9\over 2}-\zeta\right)h_2+
18\left({7\over 2}-\zeta\right)_2h_1+15\left({5\over 2}-\zeta\right)_3\, ,
\ee
that indeed depends only on $(ML)^2$ and $\zeta$.

This explicit analysis of the first three $k_n$ makes it plausible
that the propagator be \emph{analytic} in $(ML)^2$ for all spins, a
fact that we can now prove in full generality.

\subsection{General form of the current exchanges}

The general expression for the $k_n$, obtained in (\ref{sn}),
can be cast in the form
\be k_n={{\cal N}_n\over {\cal D}_n}\,  , \ee
where the denominator,
\be
{\cal D}_n=2^{3n}n!\left({5\over 2}-\zeta\right)_ng_n(\nu,\zeta)\, ,
\label{Dn} \ee
with $g_n$ defined in eq.~(\ref{gn}), is manifestly a polynomial
in $(ML)^2$ and the constant $\zeta$ was defined in
eq.~(\ref{zeta}). Notice that the poles are manifestly at the
\emph{even discrete states}. We would like to show that the numerator ${\cal N}_n$, given by
\be {\cal N}_n=2^n \sum_{r=0}^n \ \left( n \atop r \right) \ \left( \frac{1}{2} \right)_r \left({3\over
2}-\nu\right)_{n-r}\!\! (\nu-\zeta+1)_r
\left(\nu-\zeta+{3\over2}+r\right)_{n-r}\, , \label{numer} \ee
is also a polynomial in $(ML)^2$. The sum in eq.~(\ref{numer}) can
be expressed in terms of the generalized hypergeometric function
$_3F_2(a,b,c;d,e;z)$ (see Appendix A), and indeed after some
rearrangements one finds
\be {\cal N}_n=2^n\left({3\over
2}-\nu\right)_{n}\left(\nu-\zeta+{3\over2}\right)_{n} \,
_3F_2\left(-n \;,\;{1\over 2} \;, \;\nu-\zeta+1 \;;
\;\nu-n-{1\over 2} \;,\;
\nu-\zeta+{3\over 2} \;;\; 1\right)\, . \label{hyper1} \ee
Although not manifestly, this form actually defines a polynomial in
$(ML)^2$. This can be proved in several ways. To begin with, a more
symmetric expression obtains via the binomial identity (\ref{new}),
\be
\left(\nu-\zeta+{3\over2}+r\right)_{n-r}=\sum_{k=0}^{n-r} \left({n-r} \atop k \right) \ \left(\nu-\zeta+1+r\right)_{n-r-k}\left({1\over
2}\right)_k\,  , \ee
which leads to
\be {\cal N}_n=2^nn! \sum_{r+k\leq n}{1\over r!\;
k! \; (n-r-k)!}\left({1\over 2}\right)_r \left({1\over
2}\right)_k\left({3\over 2}-\nu\right)_{n-r}
(\nu-\zeta+1)_{n-k}\ .\label{nume1} \ee
Now the $\nu$-dependent sum is manifestly invariant under the
transformation
\be
\Big(\nu,\zeta\Big)\ \rightarrow\
\left(-\nu+\zeta+\frac{1}{2},\zeta\right)\, , \label{transfarg}
\ee
that interchanges the two factors $\left({3\over 2}-\nu\right)$ and
$(\nu-\zeta+1)$. Hence, it is an even polynomial in
$\left(\nu-{\zeta\over 2}-{1\over 4}\right)$, and thus a polynomial
in
\be \left(\nu-{\zeta\over 2}-{1\over 4}\right)^2=\frac{1}{4}\,
\left[\left(\zeta-{5\over 2}\right)^2-(ML)^2\right]\,, \ee
where we have made use of eq.~(\ref{masshell}). Let us also notice that
the invariance of the numerator under the above
transformation and the expression in terms of hypergeometric
functions make it possible to turn eq.~(\ref{hyper1}) into yet
another expression:
\be
{\cal N}_n=2^n\left(2-\nu\right)_{n}\left(\nu-\zeta+1\right)_{n} \
_3F_2\left(-n\; ,\ {1\over 2}\; ,\  {3\over 2}-\nu\; ;\
\zeta-\nu-n\; ,\ 2-\nu\;;\ 1\right)\, . \label{hyper2} \ee

Alternatively, one could rewrite (\ref{nume1}) in terms of the
variables $\gamma$ and $\delta$ introduced in (\ref{gd}) as
\be
{\cal N}_n=2^n \, n! \sum_{r+k\leq n} \, c_{n;\, r, k} \,
(\gamma \, - \, \delta)_{n - r} \, (\gamma \, + \, \delta)_{n - k} \, ,
\ee
where the coefficients are those of eq.~(\ref{nume1}), and read
\be
c_{n; \, r, k} \, = \, {1\over r!\;
k! \; (n-r-k)!}\left({1\over 2}\right)_r \left({1\over
2}\right)_k \, .
\ee
Next, separating in the sum the terms with $ r = k$ from those where
$r \neq k$ gives
\be \label{num}
{\cal N}_n=2^n \, n! \{
\sum_{j = 0}^{[\frac{n}{2}]} \, c_{n;\, j, j} \, h_{n - j} \, + \,
\sum_{i > j = 0}^{n} \, c_{n;\, i, j} \, h_{n - i} \,
\left[(\gamma \, + \, \delta)_{i - j} \, + \, (\gamma \, - \, \delta)_{i - j}\right] \} \, .
\ee
In order to show that this expression only involves even powers of
$(ML)^2$, we need to analyze the sum
\be
(\gamma \, + \, \delta)_{i - j} \, + \, (\gamma \, - \, \delta)_{i -
j}
\ee
that, making use of (\ref{new}), can be written as
\be
(\gamma \, + \, \delta)_{i - j} \, + \, (\gamma \, - \, \delta)_{i - j} \, = \,
\sum_{k = 0}^{i - j} \, \left( i - j \atop k \right) \, (\gamma)_{i -j -k} \,
\left[(\delta)_k \, + \, (- \delta)_k \right] \, .
\ee
Finally,  it is possible to show by induction that the quantity
\be
(\delta)_k \, + \, (- \delta)_k \,
\ee
defines an \emph{even} polynomial in $\delta$, whose explicit form
is discussed in Appendix A. As a result ${\cal N}_n$ is indeed
analytic, and actually a polynomial in $(ML)^2$ of degree $n$.

In conclusion, for all spin-$s$ totally symmetric $dS$ tensors of
mass $M$ the portions of the propagator not involving gradients,
which suffice to determine the current exchanges, take the form
\be
{\cal K}=\sum_{n=0}^{\left[ \frac{s}{2} \right]} \frac{(2-\nu)_n\;
(\nu-\zeta+1)_n}{\left(\frac{5}{2}-\zeta\right)_n\; n!}\
\frac{_3F_2\left(-n,{1\over 2},{3\over 2}-\nu;
\zeta-\nu-n,2-\nu; 1\right)}{\prod_{j=0}^{n-1}
\Big[(ML)^2+2(2j+{1})(j+3-\zeta)\Big]}\ (u^2)^n (\partial_u
\cdot \partial_u)^n J_s \, ,
\ee
so that the current exchange amplitudes are finally
\be \sum_{n=0}^{\left[ \frac{s}{2} \right]} \frac{(2-\nu)_n\;
(\nu-\zeta+1)_n}{\left(\frac{5}{2}-\zeta\right)_n\; n!}\
\frac{_3F_2\left(-n ,{1\over 2},{3\over 2}-\nu ;
\zeta-\nu-n ,2-\nu ; 1\right)}{\prod_{j=0}^{n-1}
\Big[(ML)^2+2(2j+{1})(j+3-\zeta)\Big]}\ \langle J_s^{[n]},
J_s^{[n]}\rangle \, , \label{currexfin} \ee
where the product in the denominator is simply to be read as equal
to 1 for $n=0$. This is the main result of this paper. Let us stress
that, as we have shown, this expression is actually a rational
function of $(ML)^2$, although this key property is not manifest in
this form.

A more explicit form for the propagator, and thus for the current
exchange, can be obtained computing the numerator at values of $\nu$
corresponding to the odd discrete states, that is for
$\zeta-\nu=p+1$ with an integer $p$, which defines the quantities
\be {\cal N}_n^p= 2^n\left({5\over 2}+p-\zeta\right)_{n}\left({1\over 2}-p\right)_{n} \
_3F_2\left(-n \; , \;{1\over 2}\;,\; -p\; ;\;
\zeta-p-n-{3\over 2}\; ,\; {1\over 2}-p \;;\;
1\right) \, . \ee
Using the identity principle for polynomials, ${\cal  N}_n$ can
indeed be expressed as a sum of $n$ terms, via the $n$ polynomials
obtained from $h_n$ removing one of the factors in eq.~(\ref{hn}).
More in detail, it can be recast in the form
\be {\cal N}_n=\sum_p {\cal N}_n^p\ {\prod_{p^{\;\prime}\neq
p}[(ML)^2-2p^{\;\prime}\; (2p^{\;\prime}-2\zeta+5)]\over \prod_{p^{\;\prime}\neq
p}[2p\; (2p-2\zeta+5)-2p^{\;\prime}(2p^{\;\prime}-2\zeta+5)]}\ , \ee
since both sides take the same values at the $n$ odd discrete
points. Notice that, in terms of the function $h_n$ defined in
eq.~(\ref{hn}), this expression becomes the partial-fraction
expansion
\be
{\cal N}_n=2h_n\sum_p \ {\cal N}_n^p{(-1)^p\over
(ML)^2-2p\; (2p-2\zeta+5)}{\left({5\over 2}+2p-\zeta\right)\over p\; !\; (n-p-1)! \left({5\over 2}+p-\zeta\right)_n}\, ,
\ee
which is manifestly a rational function of $(ML)^2$.

There is actually a better way of presenting our results. The idea
is to expand the numerator directly in terms of the $h_n$
polynomials of eq.~(\ref{hn}). In fact, we have found out that
identities satisfied by $_3F_2$ allow to turn ${\cal N}$ into the
expression
\be {\cal N}^{(J)}_n=\frac{1}{2^n} \, \sum_{k=0}^{n} \left( n
\atop k \right)\, \left({1\over 2}\right)_{n-k} \left({5\over
2}-\zeta+k\right)_{n-k}\ \prod_{j=0}^{k-1}
\Big[(ML)^2+2j(2j+5-2\zeta)\Big]\, . \label{nj} \ee
This form has the advantage of being manifestly expressed in terms
of $(ML)^2$, and was originally guessed by one of us after an
explicit analysis of a few cases of low $n$. In fact, eq.~(\ref{nj})
can be expressed in terms of the generalized hypergeometric function
$_3F_2$, as
\be {\cal N}^{(J)}_n=2^n\, \left( \frac{3}{2} - \nu \right)_n
\,\left(\nu - \zeta+1 \right)_n\, _3F_2\left( -n,\frac{1}{2},\zeta -
\frac{3}{2}-n;\nu-n-\frac{1}{2},\zeta-\nu-n;1 \right) \, , \ee
and coincides with ${\cal N}$ on account of the identity
\be _3F_2(-n,b,c;d,e;1)=\frac{(d-b)_n}{(d)_n}\
_3F_2(-n,b,e-c;e,b-d-n+1;1) \, ,
 \label{idhyp} \ee
that is proved in Appendix A.

These results lead to a rather compact expression for ${\cal K}$
that is also an explicitly rational function of $(ML)^2$,
\be {\cal K} = \sum_{n=0}^{\left[\frac{s}{2}\right]} \frac{(u^2)^n
J_s^{[n]}}{2^{2n}\; n! \left( \frac{5}{2}-\zeta \right)_n g_n}\
\sum_{k=0}^{n} \left( n \atop k \right)\, \left({1\over
2}\right)_{n-k}
 \left({5\over 2}-\zeta+k\right)_{n-k} h_k \, , \label{knj}
\ee
where $h_n$ and $g_n$ were defined in eqs.~(\ref{hn}) and (\ref{gn})
and the superscripts on the currents identify their successive
traces, whose first few terms read
\ba {\cal K}(x,u^a)&=&J_s+{u^2\over
4(\frac{5}{2}-\zeta)}{(ML)^2+2(\frac{5}{2}-\zeta)\over
(ML)^2-2(\zeta-3)}\ J^{\;\prime}_s\nonumber\\ &+&{(u^2)^2\over
32}{(ML)^4+8(ML)^2(\frac{7}{2}-\zeta)+12\left({5\over
2}-\zeta\right)_2
\over\left({5\over
2}-\zeta\right)_2[(ML)^2-2(\zeta-3)][(ML)^2-6(\zeta-4)]}\
J_s^{\;[2]}\nonumber\\
&+&{(u^2)^3\over 384}{(ML)^6-(ML)^4
(18\zeta-77)+92(ML)^2\left({7\over
2}-\zeta\right)_2+120\left({5\over
2}-\zeta\right)_3\over\left({5\over
2}-\zeta\right)_3[(ML)^2-2(\zeta-3)][(ML)^2-6(\zeta-4)]
[(ML)^2-10(\zeta-5)]}\ J_s^{\;[3]}\nonumber\\
&+&\dots + \left({u^2}\right)^n{{\cal N}_n\over {\cal D}_n}\
J_s^{[n]}+\dots \, ,  \label{currexp} \ea
and where in general
\be {\cal N}_n=(ML)^{2n}+\dots +{(2n)!\over
n!}\left({5\over 2}-\zeta\right)_n\, . \ee
Therefore, as we already stressed, the poles of the current-exchange
amplitudes correspond manifestly to the \emph{even} discrete states,
while the zeros are not as evident, since eq.~(\ref{nj}) involves a
sum of contributions related to the
\emph{odd} discrete states
\footnote{In order to compare with the notations of
\cite{fms}, one should make the substitution $(u^2/2)^n\rightarrow g^n
n!$, where $g$ denotes the background $dS$ metric.}.

For the sake of clarity, let us display explicitly the first few
current-exchange amplitudes fully implied by eq.~(\ref{currexp}), in
the standard notation of \cite{fms}, that for convenience are here
rescaled by an overall factor $s!$, for spin $s$, with respect to
eq.~(\ref{currexp}) :
\ba & {\bf s=1 :}  \quad  & \left(J_a \right)^2 \\
& {\bf s=2 :}  \quad  & \left( J_{ab}\right)^2  \ - \ \frac{1}{d-1}\
{(ML)^2-(d-1)\over (ML)^2-(d-2)}\ \left({J^{\;
\prime}}\right)^2 \\
& {\bf s=3 :}  \quad  & \left( J_{abc} \right)^2 \ - \
\frac{3}{d+1}\ {(ML)^2-(d+1)\over (ML)^2-d}\ \left({J^{\;
\prime}}{}_{\; a} \right)^2 \\
& {\bf s=4 :}  \quad  & \left( J_{abcd} \right)^2 \ - \
\frac{6}{d+3}\ {(ML)^2-(d+3)\over
(ML)^2-(d+2)}\ \left({J^{\; \prime}}{}_{\; ab} \right)^2 \nonumber \\
& \quad &+ \, \frac{3}{(d+1)(d+3)} \ \frac{(ML)^4 -
4(ML)^2(d+1)+3(d+1)(d+3)}{[(ML)^2-(d+2)][(ML)^2- 3 d]}\ \left({J^{\;
\prime\prime}}\right)^2 \nonumber\\
& {\bf s=5 :}  \quad  & \left( J_{abcde} \right)^2 \ - \
\frac{10}{d+5}\ {(ML)^2-(d+5)\over (ML)^2-(d+4)}\ \left({J^{\;
\prime}}{}_{\; abc} \right)^2 \\
& \quad &+ \, \frac{15}{(d+3)(d+5)} \ \frac{(ML)^4 -
4(ML)^2(d+3)+3(d+3)(d+5)}{[(ML)^2-(d+4)][(ML)^2- 3 (d+2)]}\
\left({J^{\;
\prime\prime}}{}_{\;a}\right)^2 \nonumber\\
& {\bf s=6 :}  \quad  & \left( J_{abcdef} \right)^2 \ - \
\frac{15}{d+7}\ {(ML)^2-(d+7)\over (ML)^2-(d+6)}\ \left({J^{\;
\prime}}^{\; abcd} \right)^2 \\
& \quad &+ \, \frac{45}{(d+5)(d+7)} \ \frac{(ML)^4 -
4(ML)^2(d+5)+3(d+5)(d+7)}{[(ML)^2-(d+6)][(ML)^2- 3 (d+4)]}\
\left({J^{\;
\prime\prime}}{}_{\;ab}\right)^2 \nonumber\\
& &\hspace{-2cm}- \, \frac{15[(ML)^6 -(ML)^4(9d+31)+
23(ML)^2(d+3)(d+5)-15(d+3)(d+5)(d+7)]}{(d+3)(d+5)(d+7)[(ML)^2-(d+6)][(ML)^2-
3 (d+4)][(ML)^2- 5 (d+2)]}\
\left({J^{\;
\prime\prime\prime}}\right)^2 \nonumber\\
& {\bf s=7 :}  \quad  & \left( J_{abcdefg} \right)^2 \ - \
\frac{21}{d+9}\ {(ML)^2-(d+9)\over (ML)^2-(d+8)}\ \left({J^{\;
\prime}}^{\; abcde} \right)^2 \\
& \quad &+ \, \frac{105}{(d+7)(d+9)} \ \frac{(ML)^4 -
4(ML)^2(d+7)+3(d+7)(d+9)}{[(ML)^2-(d+8)][(ML)^2- 3 (d+6)]}\
\left({J^{\;
\prime\prime}}{}_{\;  abc} \right)^2 \nonumber\\
& &\hspace{-2cm}- \, \frac{105[(ML)^6 -(ML)^4(9d+49)+
23(ML)^2(d+5)(d+7)-15(d+5)(d+7)(d+9)]}{(d+5)(d+7)(d+9)[(ML)^2-(d+8)][(ML)^2-
3 (d+6)][(ML)^2- 5 (d+4)]}\ \left({J^{\; \prime\prime\prime}}{}_a
\right)^2 \nonumber \ea

Before ending this Section, let us finally mention yet another
compact form of the propagator, that can be obtained applying the
identity (\ref{idhyp}) to the expression (\ref{hyper1}) for the
numerator. The result is also manifestly a rational function of
$(ML)^2$, but this time is expressed solely in terms of the
polynomials $g_k$ of eq.~(\ref{gn}), and reads
\be {\cal K} = \sum_{n=0}^{\left[\frac{s}{2}\right]} \frac{(u^2)^n
J_s^{[n]}}{2^{2n}\; n! \left( \frac{5}{2}-\zeta \right)_n }\
\sum_{k=0}^{n} (-1)^k\left( n \atop k \right)\, \left[\left({1\over
2}\right)_{k}\right]^2\frac{1}{g_k} \, . \ee

\section{Discussion}

In this paper we have generalized the previous massive flat-space
results of \cite{fms} to the case of $(A)dS$ backgrounds. At the
same time, we have generalized the previous $s=2$ $dS$ current
exchanges of \cite{higuchi,porrati} to the case of symmetric tensors
of arbitrary rank. Our results confirm the indications obtained for
$s=2$ in \cite{porrati}: \emph{as soon as a cosmological constant is
turned on, the current exchange amplitudes become analytic functions
of $(ML)^2$ and the vDVZ discontinuity disappears}. As we stressed
in the Introduction, however, our current grasp of higher-spin gauge
theories confines our analysis to the case of free fields, so that
we can not connect our findings with the observations of
\cite{duff}, or of \cite{porrati2}, related to the interplay of
discontinuities and interactions.

Even working with currents that are also \emph{conserved} in $d$
dimensions, as we have done in this paper, the resulting
expressions, which we presented  in their most compact form in
eq.~(\ref{knj}), display an interesting feature. Their poles
correspond in fact to
\be
(ML)^2=2(\zeta-3-j)(2j+1),\quad j=0,1,\dots , \left[{s\over
2}\right]-1 \, ,\ee
and these values identify the \emph{even discrete states} of Section
4.1, with $r=s-2-2j$. The reason for their emergence in this context
is interesting, and is related to the form of the corresponding
gauge transformations. The key observation is that the coupling to
the external current,
\be
-\int d^dx \sqrt{-g}\,  \vf \cdot J_s(x,u^a)\, ,\ee
is gauge invariant, \emph{for the even discrete states}, only if the
$\frac{(s-r)}{2}$-th trace of $J_s(x,u^a)$ vanishes. Hence, the
poles of the propagator accompanying the various traces of the
current are just like the poles that non-conserved currents would
give rise to in massless exchanges, an example of which was
displayed in eq.~(\ref{massivepole})!

As we have seen, the $dS$ current exchanges greatly simplify in the massless limit, where the numerator reduces to
\be
{\cal N}_n=2^n\left({1\over 2}\right)_n\left({5\over
2}-\zeta\right)_n\, , \ee
while the denominator reduces to
\be {\cal D}_n=2^{3n}n!\left({5\over 2}-\zeta\right)_n\left({1\over
2}\right)_n(3-\zeta)_n\, , \ee
so that $k_n$ becomes
\be
k_n={1\over n! \, 2^{2n}(3-\zeta)_n}\, . \ee
Hence, they tend smoothly to the massless result of
eq.~(\ref{fmsfin}), with \emph{no vDVZ discontinuity}.

One can similarly investigate the flat limit, that can be recovered
in the $\nu\rightarrow i \infty$ limit. In this case the dominant
term in the numerator is the $r=0$ term of the sum, since the
various contributions depend on $\nu^{2n-r}$. The denominator is
also dominated by $\nu^{2n}$, so that the ratio $k_n$ does not
depend on $\nu$ in the limit and is given by
\be k_n\rightarrow{1\over n!\,  2^{2n}\left({5\over
2}-\zeta\right)_n}\, , \ee
so that one recovers the flat massive exchanges of
eq.~(\ref{fmsfinm}.

In general, we arrived at rather compact expressions for the
spin-$s$ current exchange amplitudes in terms of generalized
hypergeometric functions of the type $_3F_2(a,b,c;d,e;z)$,
\be
\sum_{n=0}^{\left[ \frac{s}{2} \right]} \frac{(2-\nu)_n\;
(\nu-\zeta+1)_n}{\left(\frac{5}{2}-\zeta\right)_n}\
\frac{_3F_2\left(-n\; ,\ {1\over 2}\; ,\  {3\over 2}-\nu\; ,\
\zeta-\nu-n\; ,\ 2-\nu\;;\ 1\right)}{\prod_{j=0}^{n-1}
\Big[(ML)^2+2(2j+{1})(j+3-\zeta)\Big]}\ \langle J_s^{[n]}, J_s^{[n]}\rangle \,
, \label{currexfin2}
\ee
where the poles are manifest while the zeros, or the very fact that
the numerator is an analytic function of $(ML)^2$, are not. We have
also shown that the numerator is indeed a polynomial in $(ML)^2$,
and have complemented eq.~(\ref{currexfin2}) with other, manifestly
rational forms, whose zeros, however, are still not manifest. The
most compact expression of this type for the current exchange, given
in eq.~(\ref{knj}), is
\be \sum_{n=0}^{\left[\frac{s}{2}\right]} \frac{\langle
J_s^{[n]},J_s^{[n]}\rangle }{2^{2n}\; n! \left( \frac{5}{2}-\zeta
\right)_n g_n}\ \sum_{k=0}^{n} \left( n \atop k \right)\,
\left({1\over 2}\right)_{n-k}
 \left({5\over 2}-\zeta+k\right)_{n-k} h_k \, , \label{knj2}
\ee
where $h_k$ and $g_n$ are given in eqs.~(\ref{hn}) and (\ref{gn}).
The link between this result to the other expressions for ${\cal K}$
presented in Section 7 rests on an identity for the generalized
hypergeometric function $_3F_2(a,b,c;d,e;1)$, given in
eq.~(\ref{idhyp}), that is proved in Appendix A.

Unitarity is another issue of utmost importance, on which
unfortunately the current indications of our analysis are less
conclusive. In flat space, as we have stressed, unitarity is
reflected in the positivity of $J \cdot {\cal K}$ on shell. In de
Sitter space one can try to proceed along the same lines,
associating unitary exchanges to conserved currents of positive norm
and ${\cal K}$ operators that are positive definite. While the
second of these conditions is relatively easy to verify, however,
the first appears less direct in curved space. Moreover, we can now
show, with reference to the examples of spin two and three, that the
positivity of ${\cal K}$ alone is not sufficient to identify the
known regions of unitarity. Indeed, for spin two and three
\be
{\cal K}=1+k_1\ u^2 \ \partial_u \cdot \partial_u\,  , \ee
so that its eigenvectors verify the condition
\be (1-\lambda)j+k_1 \ u^2 \ \partial_u \cdot \partial_u j=0\, .
\label{unitcheck2}\ee
Applying
$\partial_u \cdot \partial_u$ to this equation while taking into account the
transversality of the current (so that in the resulting relations $d$ is
actually replaced with $d-1$) gives
\be
[1-\lambda+2k_1(d-1+2s-4)]\partial_u \cdot \partial_u j=0\,  , \ee
and one can now distinguish two possibilities. The first is that
$\partial_u \cdot \partial_u j=0$, which from
eq.~(\ref{unitcheck2}) leads to the eigenvalue $1$, while the
second is that $\partial_u \cdot
\partial_u j\neq0$, that leads to the eigenvalue
\be \lambda=1+2k_1(d+2s-5)\, . \ee
The positivity condition for the spin-2 and spin-3 propagators then
reduces to
\be 1+2k_1(d+2s-5)>0\, , \ee
and replacing in this expression $k_1$ with its explicit form of
eq.~(\ref{k1}) finally gives
\be (ML)^2>d+2s-6\, . \label{unitcheck} \ee
Under this condition, the operator ${\cal K}$ is thus positive on
conserved currents. In four dimensions and for $s=2$, the bound of
eq.~(\ref{unitcheck}) agrees precisely with the unitarity bound
obtained from the group theoretical treatment. However, for $s=3$
Group Theory gives $(ML)^2> 2d-2$, corresponding to the mass of the
second discrete state, to be contrasted with the result of
eq.~(\ref{unitcheck}), that would identify the first discrete state.
In a similar fashion, the special role of the discrete tachyonic
scalar masses that we have mentioned at the end of Section 4 is not
at all manifest in this type of analysis. Therefore, the positivity
condition for a conserved current in a de Sitter background deserves
a closer look, and we hope to return to this issue in the near
future.

\vskip 24pt


\section*{Acknowledgments}


We are very grateful to M.~Porrati for stimulating discussions at
the beginning of this project, and to Ya.S. Stanev for some
suggestions. We are also grateful to the APC-Paris VII and to the
Scuola Normale Superiore for the kind hospitality extended to one or
more of us during the course of this work. The present research was
also supported in part by INFN, by the MIUR-PRIN contract
2005-02045, by the EU contracts MRTN-CT-2004-503369 and
MRTN-CT-2004-512194, by the INTAS contract 03-51-6346, and by the
NATO grant PST.CLG.978785.

\newpage

\begin{appendix}

\section{Pochhammer symbols and hypergeometric functions}

The Pochhammer symbols $(a)_n$ are defined, for $n \in N$, as
\cite{ww}
\be
(a)_n =
\left\{
\begin{array}{cl}
1 & \mbox{if } n = 0 \\
a(a+1)\ldots (a+n-1) & \mbox{if } n \neq 0
\end{array}
\right.
 \, ,
\ee
and are simply related to the Euler $\Gamma$ function according to
\be
(a)_n = \frac{\Gamma(a+n)}{\Gamma(a)} \, .
\ee
They are a convenient device to express a number of our results, so
that, for instance
\be {(2n)!\over 2^{2n} n!}=\left({1\over
2}\right)_n \,  . \ee

A few properties of the Pochhammer symbols are used repeatedly in
the text, and are summarized below for convenience. To begin with,
one can connect rather simply the two Pochhammer symbols $(a)_n$ and
$(-a)_{n}$, according to
\be
(-a)_n = (-1)^n \, (a+1-n)_{n} \, .
\ee
In addition, the useful relation
\be
(a)_{n+k}=(a)_k (a+k)_n \label{prod}
\ee
can be proved by simply expanding both sides, while the binomial
identity
\be
\sum_{k=0}^n \left( n \atop k \right) (a)_k (b)_{n-k} = (a+b)_n  \label{new}
\ee
follows from the comparison of two ways of expanding $(1-x)^{-(a+b)}$, obtained from
\be
(1-x)^{-a} = \sum_{n=0}^\infty\, (a)_n \ \frac{x^n}{n!} \ .
\label{algpower}
\ee
The two sums
\be
\begin{split}
(a)_n \, + \, (-a)_n \, , \\
(a)_n \, - \, (-a)_n \, ,
\end{split}
\ee
can be shown by induction to define
even and odd polynomials in $a$, respectively:
\begin{align}
& (a)_{k + 1} \, + \, (- a)_{k + 1} &\,& \equiv &\,& \lambda_{2k} \, (a) &\, & = & \,
&\rho_2^{(2k)} \, a^2 \, + \, \rho_4^{(2k)} \, a^4 \, + \, \rho_6^{(2k)} \, a^6 \, + \, \dots  \, , &\\
& (a)_{k + 1} \, - \, (- a)_{k + 1} &\, & \equiv &\, & \lambda_{2k +
1} \, (a) &\, & = & \, &\rho_1^{(2k + 1)} \, a \, + \, \rho_3^{(2k +
1)} \, a^3 \, + \, \rho_5^{(2k + 1)} \,a^5 \, + \, \dots  \,  .&
\end{align}
They satisfy the system of equations
\begin{align}
&\lambda_{2k} \, (a)   \, \, \, \, \, \,   =  \, a \, \lambda_{2k-1} \, (a)\, + \, k \, \lambda_{2(k-1)} \, (a)\, ,\\
&\lambda_{2k+1} \, (a)  =  \, a \, \lambda_{2(k-1)} \, (a)\, + \, k \, \lambda_{2k-1} \, (a)\, ,
\end{align}
whose solution can be found using the identity principle for
polynomials. For the first few cases
\be
\begin{split}
&\rho_1^{(2k + 1)} \, = \, 2 \, k! \, \, , \\
&\rho_2^{(2k)} \, = \, 2 \, k! \, \sum_{i = 1}^{k} \, \frac{1}{i} \, , \\
&\rho_3^{(2k + 1)} \, = \, 2 \, k! \, \sum_{i < j = 1}^{k} \, \frac{1}{i \cdot j} \, , \\
&\rho_4^{(2k)} \, = \, 2 \, k! \, \sum_{i_1 < i_2 < i_3 = 1}^{k} \, \frac{1}{i_1 \cdot i_2 \cdot i_3} \, ,
\end{split}
\ee
while in general, for both $\lambda_{2k}$ and $\lambda_{2k+1}$
\be \rho_{l} \, = \, 2 \, k! \, \sum_{i_1 < \dots < i_{l} = 1}^{k}
\, \frac{1}{i_1 \cdot . . . \cdot i_{l}} \, . \ee

The identity \cite{ww}
\be
\sum_{n=0}^r \left( r \atop n \right) (-1)^n \frac{(a)_n}{(b)_n} = \frac{(b-a)_r}{(b)_r} \, \label{gauss}
\ee
follows from the Gauss relation
\be
_2F_1(a,b;c;1) = \frac{\Gamma(c) \Gamma(c-a-b)}{\Gamma(c-a)
\Gamma(c-b)} \, ,\label{gauss2}
\ee
that in its turn can be derived, up to an analytic continuation,
from the Gauss integral
\be
_2F_1(a,b;c;z)= \frac{\Gamma(c)}{\Gamma(b) \; \Gamma(c-b)}\ \int_0^1 dt \
t^{b-1} \; (1-t)^{c-b-1} \; (1-zt)^{-a}
\ee
for the hypergeometric function $_2F_1(a,b;c;z)$, that is valid for
$Re(c)>Re(b)>0$. It is interesting to recall that this key result
follows inserting eq.~(\ref{algpower}) in the series expansion
\be
_2F_1\big(a,b;c;z\big)= \sum_{n=0}^\infty \frac{(a)_n (b)_n}{(c)_n}
\ \frac{z^n}{n!} \, .
\ee
defined for $|z|<1$, which turns the sum into
\be
_2F_1\big(a,b;c;z\big)\, = \, \frac{\Gamma(c)}{\Gamma(b)\Gamma(c-b)}
\,  \int_0^1 dt \sum_{n=0}^\infty \frac{t^{b+n-1}\; (a)_n \; z^n}{n!} \ (1-t)^{c-b-1} \, .
\ee

In a similar fashion, the generalized hypergeometric function
$_3F_2\big(a,b,c;d,e;z\big)$ can be defined, for $|z|<1$, via the
series expansion
\be
_3F_2\big(a,b,c;d,e;z\big)= \sum_{n=0}^\infty \frac{(a)_n (b)_n
(c)_n}{(d)_n (e)_n} \ \frac{z^n}{n!} \, ,
\ee
and similar steps lead to the integral representation
\be
_3F_2(a,b,c;d,e;z)= \frac{\Gamma(e)}{\Gamma(c)\Gamma(e-c)} \
\int_0^1 dt \; t^{c-1} \, (1-t)^{e-c-1} \; {_2F_1}(a,b;d;tz) \, .
\label{3f2int}
\ee

Here we are particularly interested in the special case $a=-n$,
where hypergeometric series collapse to finite sums, so that, in
particular,
\be
_2F_1(-n,b;c;z) = \sum_{k=0}^n (-1)^k
\left(n \atop k \right) \, \frac{(b)_k}{(c)_k} \ z^k\, ,
\ee
\be
_3F_2(-n,b,c;d,e;z) = \sum_{k=0}^n (-1)^k
\left(n \atop k \right) \, \frac{(b)_k (c)_k}{(d)_k (e)_k} \ z^k\, .
\ee

Eq.~(\ref{idhyp}) follows if the integral representation
(\ref{3f2int}) is combined with the relation
\be
_2F_1(-n,b;d;t)=\frac{(d-b)_n}{(d)_n}\ _2F_1(-n,b;b-d+1-n;1-t) \, ,
\label{id2f1}
\ee
which leads to
\be
_3F_2(-n,b,c;d,e;z)=
\frac{(d-b)_n\Gamma(e)}{(d-b)_n (d)_n\Gamma(c)\Gamma(e-c)}
\
\int_0^1 dt \; t^{e-c-1} \, (1-t)^{c-1} \; {_2F_1}(-n,b;b-d+1-n;t) \,
,
\nonumber
\ee
after the integration variable is redefined interchanging $t$ and
$1-t$. Notice indeed that, up to the two Pochhammer symbols, this
last expression can be obtained from eq.~(\ref{3f2int}) letting
\ba
&c \ \to \ e - c \, ; \qquad
&e \ \to \ e \, ; \nonumber \\
&b \ \to \ b \, ; \qquad &d \ \to \ b-d+1 - n \, .
\ea
so that the end result is indeed eq.~(\ref{idhyp}), albeit in the
equivalent form
\be _3F_2(-n,b,c;d,e;1)=\frac{(d-b)_n}{(d)_n}\
_3F_2(-n,b,e-c;b-d-n+1,e;1) \, ,
 \label{idhyp2} \ee
with the two (symmetric) lower arguments interchanged.
Eq.~(\ref{id2f1}) is a special case of the basic continuation
formula
\ba
{_2F_1}(a,b;c;z)=\!\!&&\frac{\Gamma(c)\Gamma(a+b-c)}{\Gamma(a)\Gamma(b)}
\ (1-z)^{c-a-b} \ {_2F_1}(c-a,c-b;c-a-b+1;1-z)\nonumber \\
+&&\frac{\Gamma(c)\Gamma(c-a-b)}{\Gamma(c-a)\Gamma(c-b)} \
{_2F_1}\left(a,b;a+b+1-c;1-z\right) \, , \ea
that follows from Barnes' lemma, the contour integral representation
\be \frac{1}{2\pi i} \ \int_{-i \infty}^{i \infty}
\frac{\Gamma(a+s)\Gamma(b+s)}{\Gamma(c+s)}\ \Gamma(-s)\, (-z)^{s}
\, ds = \frac{\Gamma(a)\Gamma(b)}{\Gamma(c)}\ {_2F_1}(a,b;c;z) \,
,\label{barnes} \ee
and is discussed, for instance, in \cite{ww}.

\section{Proof of eq.~(\ref{fff})}

In this Appendix we would like to provide some details on the
derivation of eq.~(\ref{fff}), since similar manipulations are used
repeatedly in the main body of this paper. The starting point is
eq.~(\ref{fjs}), in which one expands $(u^2+v^2)^n$ using Newton's
binomial identity, obtaining
\be
f=\sum_{m=0}^{[\frac{s}{2}]} \
\sum_{p=0}^{[\frac{s}{2}]-m} \ \sum_{n=0}^{[\frac{s}{2}]-p-m}
{{(v^2)^{n+p} (u^2)^m (-1)^p}\over{2^{2(n+p+m)}\; p!\; m!\; n!}}\
\frac{\left({3\over 2}-\nu\right)_{n+m}}{\left({5\over
2}-\zeta\right)_{n+m}\, (2-\nu)_{p+n+m}}\ (\partial_u \cdot
\partial_u)^{p+n+m}\, J_s \, .
\label{fjs2} \ee
In order to arrive at this expression from eq.(\ref{fjs}), one actually needs to reorder the summations according to
\be
\sum_{n=0}^{[\frac{s}{2}]}\sum_{p=0}^{[\frac{s}{2}]-n}\sum_{m=0}^{n} = \sum_{p=0}^{[\frac{s}{2}]}\sum_{n=0}^{[\frac{s}{2}]-p}\sum_{m=0}^{n} =
\sum_{p=0}^{[\frac{s}{2}]}\sum_{m=0}^{[\frac{s}{2}]-p}\sum_{n=m}^{[\frac{s}{2}]-p}=
\sum_{m=0}^{[\frac{s}{2}]}\sum_{p=0}^{[\frac{s}{2}]-m}\sum_{n=m}^{[\frac{s}{2}]-p}= \sum_{0\leq m+n^\prime+p \leq [\frac{s}{2}]}\, ,
\ee
where $n^\prime=n-k$. One can then insert in the sum the factor
\be
\sum_{r=0}^{[\frac{s}{2}]} \delta_{r,n^\prime+p}\, ,
\ee
which is unity on account of the fact that, in the complete
summation, $0\leq n^\prime+p \leq [\frac{s}{2}]$, so that one of the
Kronecker $\delta$'s is always effective. The order of the resulting
$p$ and $n^\prime$ summations can now be inverted, according to
\be
\sum_{m=0}^{[\frac{s}{2}]} \sum_{p=0}^{[\frac{s}{2}]-k}\sum_{n^\prime=0}^{[\frac{s}{2}]-p-k}\sum_{r=0}^M =
\sum_{m=0}^{[\frac{s}{2}]} \sum_{r=0}^M \sum_{n^\prime=0}^{[\frac{s}{2}]-m}\sum_{p=0}^{[\frac{s}{2}]-n^\prime-m},
\ee
in order to begin from the summation over $p$. This amounts to the
replacement, induced by the $\delta$'s, of $p$ with $r-n^\prime$,
which also lowers the upper bound on $n^\prime$ to $r$, so that
\be
f=\sum_{m=0}^{[\frac{s}{2}]} \
\sum_{p=0}^{[\frac{s}{2}]-m} \ \sum_{n=0}^{[\frac{s}{2}]-p-m}
{{(v^2)^{n+p} (u^2)^m (-1)^p}\over{2^{2(n+p+m)}\; p!\; m!\; n!}}\
\frac{\left({3\over 2}-\nu\right)_{n+m}}{\left({5\over
2}-\zeta\right)_{n+m}\, (2-\nu)_{p+n+m}}\ (\partial_u \cdot
\partial_u)^{p+n+m}\, J_s \, .
\label{fjs2} \ee
Another identity needed to arrive at the final expressions of
eqs.~(\ref{fff}) and (\ref{bb}) is eq.~(\ref{prod}). Finally, the
pair of summations over $m$ and $r$ can effectively be cut to the
region $m+r \leq [\frac{s}{2}]$, since for larger values the
corresponding traces of the current do not exist, and the
differential operator $(\partial_u \cdot \partial_u)^{m+r}$
annihilates $J_s(x,u^a)$.

\newpage

\section{Proof of eq.~(\ref{prop1})}

In this Appendix we would like to provide a proof of the expression (\ref{prop1}) for the
propagator. In order to simplify the notation, as in Section 5 it is convenient to define
\be a=\mu-s=2(\nu-\zeta)\, .\ee
In addition, let us recall that
\ba
E_0(t)&=&(1+t)(2+t)(3+t)(a+2+t)(a+3+t)(a+4+t)\, , \nonumber\\
E_2(t)&=&3(1+t)[t(a+2+t)^2+(a+4+t)(a+2+t)]\, , \nonumber\\
E_4(t)&=&3t(a+1+t)+6\, , \nonumber \\
E_6(t)&=&1 \, ,
\ea
The first step in the proof is the explicit computation of the first four $\lambda_i$, for
$i=0,\dots 4$, for which eq.~(\ref{kappa}) takes a special form.  The result,
\ba &\lambda_0&=1\, ,\nonumber \\
&\lambda_1&=-{1\over 2(a+3)}\, ,\nonumber \\
&\lambda_2&={1\over 8(a+3)(a+5)}\, , \nonumber \\
& \lambda_3&=-{1\over 48(a+3)(a+5)(a+7)},\nonumber \\
&\lambda_4&={1\over 384(a+3)(a+5)(a+7)(a+9)}\, ,
\label{firstlambda}\ea
displays a nice pattern, with poles at odd negative values of $a$, in agreement with eq.~(\ref{lam}).
For $n>4$, eq. (\ref{kappa}) simplifies, since only the ${\cal K}_i$ contribute, and
\ba
\lambda_n&=&3(a+4)\sum_{i_1+\dots i_p=n-2}(-1)^p
{E_{2i_1}(1+2i_1)E_{2i_2}(1+2i_1+2i_2)\dots E_{2i_p}(2n-3)\over
E_0(1)E_0(1+2i_1)E_0(1+2i_1+2i_2)\dots E_0(2n-3)} \nonumber\\
&+&{7a+36\over2(a+3)}\sum_{i_1+\dots i_p=n-3}(-1)^p
{E_{2i_1}(3+2i_1)E_{2i_2}(3+2i_1+2i_2)\dots E_{2i_p}(2n-3)\over
E_0(3)E_0(3+2i_1)E_0(3+2i_1+2i_2)\dots E_0(2n-3)}\nonumber\\
&+&{1\over 2(a+3)}\sum_{i_1+\dots i_p=n-4}(-1)^p
{E_{2i_1}(5+2i_1)E_{2i_2}(5+2i_1+2i_2)\dots E_{2i_p}(2n-3)\over
E_0(5)E_0(5+2i_1)E_0(5+2i_1+2i_2)\dots E_0(2n-3)}\, .
\ea
These are still rather cumbersome expressions, but performing explicitly the sums over $i_p=1,2,3$, one can derive
the relatively handy difference equation
\be
\lambda_n=-{1\over E_0(2n-3)}\ \Big[
E_2(2n-3)\lambda_{n-1}+E_4(2n-3)\lambda_{n-2}+
\lambda_{n-3}\Big]\, .\label{rec} \ee

A proof of eq.~(\ref{lam}) can now be obtained by induction,
starting from the observation that the same recursion relation holds
for the first four coefficients in eq.~(\ref{firstlambda}), that can
also serve as initial conditions. On the other hand, the relation
underlying eq.~(\ref{lam}) is
\be \lambda_m=- \frac{\lambda_{m-1}}{2m(a+2m+1)} \ ,
\label{lambdaothers}\ee
that as we have stressed is also a special solution of (\ref{rec}).
Thus, if we assume that eq.~(\ref{lambdaothers}) hold for $m=n-1$,
eq.~(\ref{rec}) implies that it also does for $m=n$.
\end{appendix}

\end{document}